\documentclass[aps,amsmath,amssymb,nofootinbib,twocolumn,pr]{revtex4-2}%
\usepackage{graphicx}
\usepackage{color}
\usepackage{verbatim} 
\usepackage{color}
\usepackage{bbold}
\usepackage{dcolumn}
\usepackage{bm}
\usepackage{upgreek}
\usepackage[unicode=true,colorlinks=true,citecolor=blue,urlcolor=blue]{hyperref}
\usepackage[normalem]{ulem}
\usepackage{physics}

\newcommand*\xoverline[2][0.75]

\begin{document}

\title{
%Energy and momentum relaxation rates determine optical absorption and photocurrents in topological insulators at high intensity\\
Nonlinear optical absorption and photocurrents in topological insulators} 
%governed~by~elastic scattering}

\author{N.~V.~Leppenen}
\author{L.~E.~Golub}	
\affiliation{Ioffe Institute, 194021 St. Petersburg, Russia}

%\date{\today}

\begin{abstract}
Theory of optical absorption of linearly and circularly polarized light in surface and edge states in topological insulators is developed for a nonlinear in light intensity regime. The absorbance for surface states and the absorption width for the edge states both decease as $1/\sqrt{I}$ at high light intensity $I$. Elastic scattering of the photoexcited electrons and holes is taken into account. The absorption bleaching at high intensity is shown to be strongly suppressed by the elastic scattering. The linear-circular dichroism of one- and two-photon absorption in surface states, emerging in the nonlinear in intensity regime, decreases due to the elastic scattering of photocarriers. 
Absorption in the edge states of two-dimensional topological insulators also depends on the elastic scattering rate being suppressed when it is stronger that the energy relaxation rate. 
The linear dichroism -- a dependence of the absorption length on the polarization plane orientation -- is studied at arbitrary light intensities. We show that a degree of the linear dichroism is governed by the ratio of the elastic and inelastic relaxation times.
The photocurrents generated in the edge states by both linearly and circularly polarized light change their intensity dependencies from the linear one at low $I$ to the $\propto \sqrt{I}$ at high intensities. The variation of the photocurrents with both polarization and intensity are strongly dependent on the elastic scattering rate. The considered effects can be observed in THz frequency range at laser intensities used in modern experiments.
\end{abstract}

\maketitle

\section{Introduction}

Topological insulators are very important due to existence of topologically protected states localized at the sample boundaries with energy in the bulk energy gap. They are two-dimensional (2D) surface states and 1D edge states in 3D and 2D topological insulators, respectively~\cite{TI_book}.
Time-reversal symmetry and spin-momentum locking result in the absence of backscattering of these states. While for 2D surface states this property results in some important consequences in transport,  for 1D edge this makes the transport totally ballistic because the backscattering is an only elastic relaxation channel.
However, the elastic scattering times of 1D edge states are reported to be as short as tens of picoseconds~\cite{Gusev_Kvon_2019}. The sources of such an effective scattering are actively debated: they are magnetic impurities~\cite{TI_book}, 
%charge puddles~\cite{scatt_puddles}, 
nonmagnetic defects in the presence of interaction~\cite{scatt_nonmagn_imp_interaction1,scatt_nonmagn_imp_interaction2,scatt_nonmagn_imp_interaction3}, etc.

While transport measurements give an access to momentum scattering time of charge carriers, a low-intensity optical absorption is insensitive to the carrier kinetics. However, in the nonlinear in  intensity regime, relaxation processes affect light absorption. For interband transitions in semiconductors, various relaxation processes limit the absorption~\cite{Parshin_Shabaev,Rasulov,
Agarwal2017}. 
%For graphene, the product of the longitudinal and transverse relaxation rates is the parameter determining the absorption saturation~\cite{Agarwal2017}. 
The main relaxation channel important for absorption bleaching is the energy relaxation by phonons. Studies of energy relaxation mechanisms in topological insulators  reveal that it occurs via emission of phonons in charge puddles~\cite{scatt_puddles,en_rel_Kvon}.

%While low-intensity optical and transport measurements give an access to momentum scattering time of charge carriers, the energy relaxation time is more subtle. 
%
%The relaxation times do not affect light absorption at low intensity but they determine the absorption if the intensity is high. The relaxation times in the initial and final states determine the bleaching and saturation intensity for direct interlevel transitions~\cite{Boyd_book}. 
%For interband transitions in semiconductors, various relaxation processes limit the absorption~\cite{Parshin_Shabaev,Rasulov}. The main relaxation channel important for absorption bleaching is the energy relaxation by phonons. For graphene, the product of the longitudinal and transverse relaxation rates is the parameter determining the absorption saturation~\cite{Agarwal2017}.

One can see that the transport and optical experimental results give information about momentum and energy relaxation rates, respectively. However, the situation becomes more interesting if the momentum scattering in the final states of direct optical transitions is effective. In this case, the absorbed power is determined by a competition of energy and momentum relaxation processes depending strongly on the ratio of corresponding times. It has been shown that studies of high-power optical absorption in 3D Weyl semimetals makes allowance for investigating both energy and momentum relaxation processes in optical experiments~\cite{pssb_2019}.

%An important step further is account for scattering in the final states of direct optical transition.
%Two independent parameters: the dimensionless electric field amplitude and the ratio of elastic and inelastic scattering rates~\cite{pssb_2019}. We have shown that the high-power optical absorption in 3D Weyl semimetals is sensitive to the ratio of the momentum and energy relaxation times which makes allowance for investigating relaxation processes from optical experiments. 

%However we have shown that the high-power optical absorption and photocurrent in 3D Weyl semimetals is sensitive to the ratio of the momentum and energy relaxation times which makes allowance for investigating relaxation processes from optical and photogalvanic experiments. 

Moreover, light absorption in non-centrosymmetric systems is accompanied by generation of photocurrents. Usually, the photogalvanic currents are linear in the light intensity. However, at high intensity the photocurrent saturates, and study of the photocurrent saturation gives information on various relaxation processes of carriers~\cite{Bleaching_holes,Dora2012,Artemenko2013,ArtemenkoPRB,Dantas2021,KovalevEntin2021,edge_nonlinear,BiTe_exp}. It has been shown that the photocurrent saturation in 3D Weyl semimetals is also governed by both energy and momentum relaxation rates of photocarriers~\cite{pssb_2019}.

In the present work, we develop a theory of light absorption and photogalvanic effects for direct optical transitions between the topologically protected states
%between the surface states and between the edge states 
%in topological insulators 
at arbitrary light intensities. The main feature is that elastic momentum scattering of the photocarriers is taken into account.
We derive the intensity dependence of the absorption and the photocurrent and  show that corresponding  measurements allow determination of the momentum and energy relaxation rates of the surface and edge states. 

The paper is organized as follows. In Sec.~\ref{General} we derive general expressions for the nonequilibrium occupations and absorption. In Sec.~\ref{2DSS} this approach is used for calculation of the absorbance 
in 2D topological insulators. Section~\ref{1DES} is devoted to absorption and photocurrent at transitions between edge states in 1D topological insulators. The results are discussed in Sec.~\ref{Disc}.
Concluding remarks are given in Sec~\ref{Concl}.

\section{General theory}
\label{General}

\begin{figure}[t]
\centering
\includegraphics[scale=0.45]{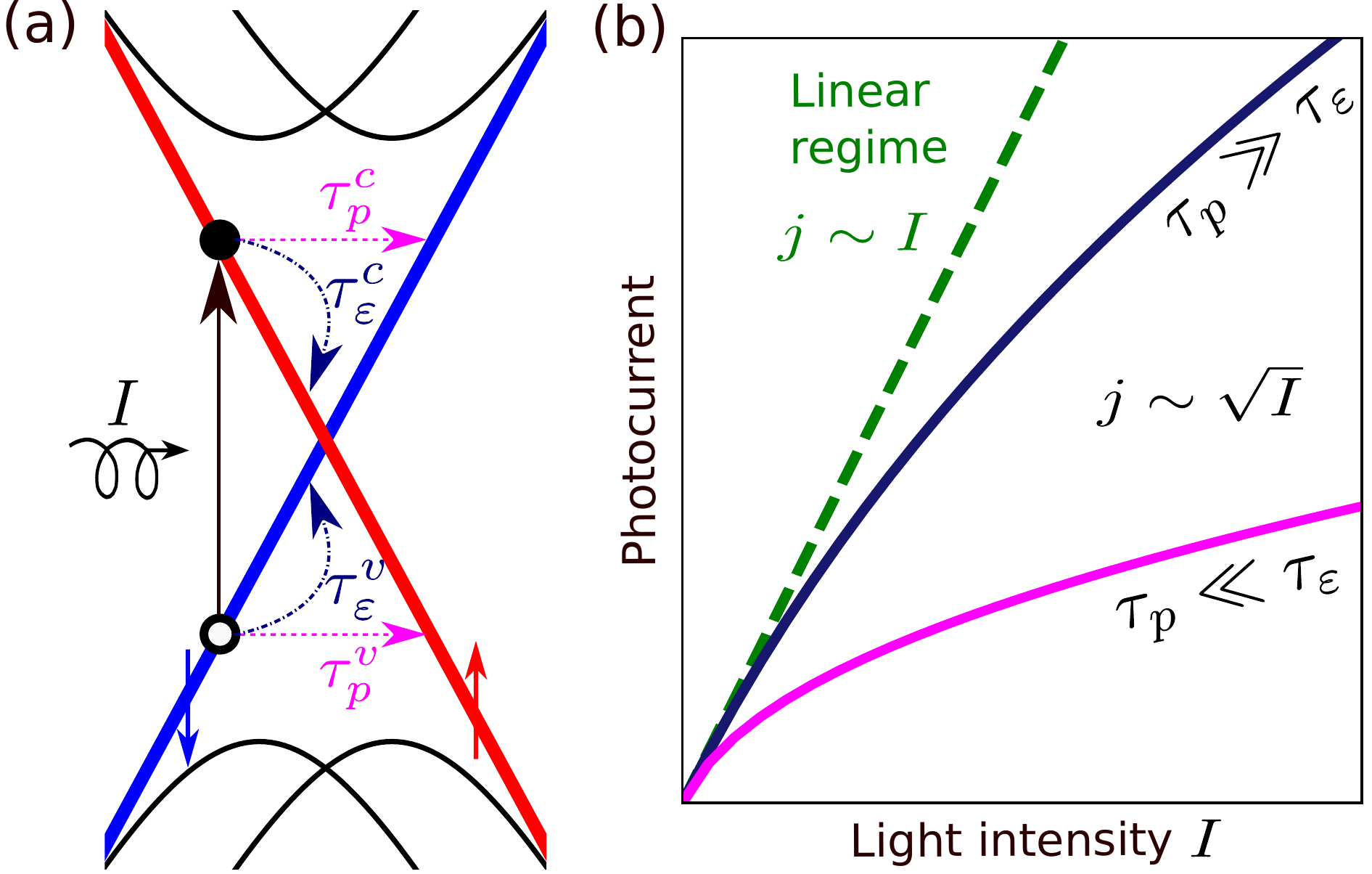} 
\caption{ (a) Scheme of direct optical transitions between topologically protected edge states. The photoexcited carriers 
%in the final state experience energy and momentum relaxation processes 
relax by energy and momentum
with the characteristic times $\tau_\varepsilon$ and $\tau_p$, respectively. The black curves represent energy  dispersion in the bulk with the account for the spin-orbit splitting.
%interface coupling~\cite{Zunger_gamma}.   
(b) Sketch of the photocurrent dependence on the light intensity. The dashed line shows the linear in the intensity photocurrent. At high intensity the current increases as $j \propto \sqrt{I}$ with the prefactor dependent on 
the relation between $\tau_\varepsilon$ and $\tau_p$. }
\label{fig:sketch}
\end{figure}

We consider direct optical transitions between the valence ($v$) and conduction ($c$) bands of the topologically protected states.
We use the kinetic equation approach for description of nonlinear optical absorption.
In the steady state, the electron occupations $f_{c,v}$ 
satisfy the following system of kinetic equations:
\begin{subequations}
\label{sys1}
\begin{align}
&{f_c-f_c^0\over \tau_\varepsilon^c} + {f_c-\left<f_c\right>\over \tau_p^c} = G(f_v-f_c),\\
&{f_v-f_v^0\over \tau_\varepsilon^v} + {f_v-\left<f_v\right>\over \tau_p^v} = -G(f_v-f_c).
\end{align}
\end{subequations}
Here $\tau_p^i$ and $\tau_\varepsilon^i$ ($i=c,v$) are the momentum and energy relaxation times in the bands, Fig.~\ref{fig:sketch}(a), $G$ is the generation rate of the direct optical transition, angular brackets mean averaging over directions of the wavevector at its fixed absolute value, and $f_i^0$ are the equilibrium distributions.
The first and second terms in the left-hand sides of Eqs.~\eqref{sys1} describe inelastic relaxation of the nonequilibrium distributions to $f_{c,v}^0$  and elastic scattering, respectively.

We introduce the nonequilibrium corrections $\Delta f_i$ as $f_{c,v}=f_{c,v}^0 \pm \Delta f_{c,v}$ and rewrite Eqs.~\eqref{sys1} as follows
% Equations for them have the form
\begin{equation}
\label{kin_eq_i}
{\Delta f_i \over \tau_i} - {\left<\Delta f_i\right>\over \tau_p^i} = G\qty(\mathcal{F}-\Delta f_c - \Delta f_v).
\end{equation}
Here $\mathcal{F}=f_v^0-f_c^0$ is the difference of the occupations in equilibrium, and the total relaxation rates in the bands are defined as
\begin{equation}
{1\over \tau_i} = {1\over \tau_p^i} + {1\over \tau_\varepsilon^i}.
\end{equation}
We also introduce a sum and a difference of the occupation corrections
\begin{equation}
\Delta f_\pm = \Delta f_c \pm \Delta f_v.
\end{equation}
Taking a sum  and a difference of the kinetic Eqs.~\eqref{kin_eq_i} (with $i=c,v$) we obtain equations for $\Delta f_\pm$. 
The equation for $\Delta f_-$ is homogeneous, and its averaging yields
a linear relation between $\left<\Delta f_+\right>$ and $\left<\Delta f_-\right>$. Then we exclude $\Delta f_-$ and obtain an equation containing only $\Delta f_+$ and $\left<\Delta f_+\right>$. Averaging it over directions of the wavevector we find $\left<\Delta f_+\right>$, and then finally get $\Delta f_+$.
The result is:
\begin{equation}
\label{fcv1}
f_v-f_c = {\mathcal{F}\over (1+2G\bar{\tau})\qty[1+\Psi\qty(\bar{\tau}_\varepsilon/\bar{\tau}-1)]}.
\end{equation}
Here we took into account that $\Delta f_+=\mathcal{F}-(f_v-f_c)$, and introduced the mean relaxation times
\begin{equation}
\bar{\tau} = {\tau_c + \tau_v\over 2}, \qquad \bar{\tau}_\varepsilon = {\tau_\varepsilon^c + \tau_\varepsilon^v\over 2},
\end{equation}
and the function $\Psi$ dependent on the absolute value of the wavevector:
\begin{equation}
\label{Psi_def}
\Psi = \left< {G\over G + 1/(2\bar{\tau})}\right>.
\end{equation}
The obtained expressions generalize the results of Ref.~\cite{pssb_2019} to the case of asymmetric conduction and valence bands with different relaxation times.

The two factors in the denominator of Eq.~\eqref{fcv1} describe decrease of the occupation difference and, hence, of the optical transition probability with increase of the light intensity. The first is a usual bleaching of absorption depending on the total relaxation time while the second depends on the ratio of the energy and momentum relaxation rates. It describes an additional effect caused by momentum scattering in the final states.

The absorbed power at direct optical transitions between the conduction and valence bands is given by
\begin{equation}
{\alpha I \over \hbar \omega}
=\sum_{\bm k}G(f_v-f_c)
% = {\cal F} \sum_{\nu,\bm k}{\Psi_k/(2\tau)\over 1 + \Psi_k \tau_\varepsilon/\tau_p}.
 = {{\cal F} \over 2\bar{\tau}}\sum_{\bm k}{\Psi \over 1+\Psi\qty(\bar{\tau}_\varepsilon/\bar{\tau}-1)}.
\end{equation}
Here $\alpha$ is the absorbance or the absorption length in 2D or 1D case, respectively, and $I$ and $\omega$ are the intensity and frequency of radiation.

The interband generation rate has the following form
\begin{equation}
\label{G_k}
G = {2|M_{cv}(\bm k)|^2/\tau \over [\varepsilon_c(k)-\varepsilon_v(k)-\hbar\omega]^2 + (\hbar/\tau)^2},
\end{equation}
where $M_{cv}(\bm k)$ is the matrix element of the direct optical transition, $\bm k$ is the wavevector of the initial and final states in the conduction and valence bands, $\varepsilon_{c,v}$ are the conduction- and valence-band dispersions, and the relaxation rate  $1/\tau$ is  a
half-sum of the total relaxation rates in the bands: 
\begin{equation}
{1\over \tau} = {1\over 2}\qty({1\over \tau_c}+{1\over \tau_v}).
\end{equation}
Both the generation rate and the function $\Psi$ have sharp maxima at $k=k_\omega$ where $k_\omega$ is determined by the energy conservation: $\varepsilon_c(k_\omega)-\varepsilon_v(k_\omega)=\hbar\omega$. Therefore we put $k=k_\omega$ everywhere except for the denominator in Eq.~\eqref{G_k}, introduce $\Delta=[\varepsilon_c(k)-\varepsilon_v(k)-\hbar\omega]\tau/\hbar$ and assuming $\omega \tau \gg 1$  obtain:
\begin{equation}
\label{alpha}
{\alpha\over \alpha_0} = {2\tau\over \pi {\cal E}^2 \bar{\tau}}\int\limits_0^\infty d\Delta {\Psi\over 1+ \Psi\qty(\bar{\tau}_\varepsilon/\bar{\tau}-1) }.
\end{equation}
Here 
$\alpha_0$ is the low-intensity value of $\alpha$ independent of the light intensity and relaxation times,
%$\Psi(\Delta)$ is given by  Eq.~\eqref{Psi},
the dependence $\Psi(\Delta)$
%, 
follows from Eqs.~\eqref{Psi_def},~\eqref{G_k}:
%is given by
\begin{equation}
\label{Psi}
\Psi(\Delta) = \left< { |2\tau M_{cv}(\bm k)/\hbar|^2\over  (1 + \Delta^2)\tau/\bar{\tau} + |2\tau M_{cv}(\bm k)/\hbar|^2}\right>,
\end{equation}
where averaging is performed at $k=k_\omega$, and the dimensionless electric field amplitude  defined as
\begin{equation}
\label{E}
{\cal E} = {2\tau\over \hbar}\sqrt{\left< |M_{cv}(\bm k)|^2\right>},
\end{equation}
%Note that ${\cal E}$ 
is a product of the Rabi frequency and the relaxation time.

Below we apply this general formalism to direct optical transitions between the topologically protected states 
in 3D and 2D topological insulators. It will be demonstrated that absorption and photocurrent in these systems saturate at high intensity as $\alpha \propto 1/\sqrt{I}$ and $j \propto \sqrt{I}$ 
%at high intensity and 
with the prefactors strongly dependent on the ratio of the relaxation times, Fig.~\ref{fig:sketch}(b).

%If the symmetry of the system allows for generation of photocurrents then the photocurrent density is given by
%\begin{equation}
%\bm j = e \sum_{\bm k} (\bm v_c f_c + \bm v_v f_v),
%\end{equation}
%where $\bm v_{c,v}=\hbar^{-1}\partial \varepsilon_{c,v}/\partial \bm k$ are the velocities in the bands.
%Then, from the expression for nonequilibrium occupations~\eqref{fcv1},~\eqref{fcv2} we obtain:
%\begin{equation}
%\bm j = {e \mathcal{F}\over 2\bar{\tau}}\sum_{\bm k}
%{\bm \Phi_c \tau_c - \bm \Phi_v \tau_v\over 1+\Psi\qty(\bar{\tau}_\varepsilon/\bar{\tau}-1)},
%\end{equation}
%where we introduced the vectors
%\begin{equation}
%\bm \Phi_{c,v} = \left< \bm v_{c,v}{G\over G + 1/2\bar{\tau}}\right>.
%\end{equation}

\section{2D surface states}
\label{2DSS}

%\subsection{Nonlinear absorption at direct band to band optical transitions}

Conduction and valence surface bands in topological insulators are described by the linear in momentum effective Hamiltonian
\begin{equation}
\label{H}
{\cal H} = \hbar v_0 [\bm \sigma \times \bm k]_z,
\end{equation}
where $z$ is a direction normal to the surface,  $\sigma_{x,y}$ are Pauli matrices, and $v_0$ is the surface state's velocity. The energy dispersions $\varepsilon_c(k)$ and $\varepsilon_v(k)$ are symmetric in this model: ${\varepsilon_c=-\varepsilon_v=\hbar v_0 k}$, therefore $k_\omega = \omega/(2v_0)$.
%, Fig.~\ref{fig:sketch}(a) \commentNL{Maybe only write that we have linear spectrum here, because Fig 1 (a) shows edge states? Or another figure.. For example, add inset on 2(a).}.
The direct optical transition matrix element is given by
\begin{equation}
\label{M_cv}
M_{cv}(\bm k) = {ev_0 \over \omega}{[\bm E_0 \times \bm k]_z \over k},
%{{\cal E} \hbar \over \sqrt{2}\tau} {[\bm e \times \bm k]_z \over k}
\end{equation}
where $\bm E_0$ is the complex amplitude of the radiation electric field
$\bm E(t) = \bm E_0 \exp(-i\omega t)+c.c.$
This yields for normally incident radiation
\begin{equation}
\label{M_cv_quad}
{\cal E} = \sqrt{2}{ev_0\tau E_0 \over \hbar \omega}, \quad
|M_{cv}(\bm k)|^2 = \qty({{\cal E} \hbar \over 2\tau})^2 (1-P_\text{lin}\cos{2\varphi}).
\end{equation}
Here 
%$P_L=\sqrt{1-P_\text{circ}^2}$ 
$P_\text{lin}$ is the linear polarization degree, $E_0 \equiv \abs{\bm E_0}$,
and $\varphi$ is the angle between $\bm k$ and the direction of the preferred linear polarization. The low-intensity absorbance is polarization-independent and is given by the fine-structure constant:
\begin{equation}
\eta_0 = {\pi e^2\over 4\hbar c} \mathcal{F},  
\end{equation}
where 
\begin{equation}
\label{F}
\mathcal{F}=f_0(-\hbar\omega/2)-f_0(\hbar\omega/2)
\end{equation}
with $f_0(\varepsilon)$ being the Fermi-Dirac distribution.
The electron-hole symmetry also implies that the relaxation rates in the bands are equal: 
$\tau_{\varepsilon,p}^{c}=\tau_{\varepsilon,p}^{v}\equiv \tau_{\varepsilon,p}$,
%$\tau_c=\tau_v=\tau$, 
so the total relaxation rate reads
\begin{equation}
\label{inverse_tau}
{1\over \tau} = {1\over \tau_\varepsilon}+{1\over \tau_p}.
\end{equation}

\subsection{One-photon absorption}

It follows from Eq.~\eqref{M_cv_quad} that the matrix element squared is angular-independent at circular polarization but it has a dependence on the orientation of $\bm k$ at linear polarization. This results in a difference in the functions $\Psi$ for these two cases:	
\begin{subequations}
\label{Psi_1}
\begin{equation}
\label{Psi_1_circ}
\Psi_\text{circ} = {{\cal E}^2  \over 1+{\cal E}^2+\Delta^2},
\end{equation}
\begin{equation}
\label{Psi_1_lin}
\Psi_\text{lin} = 1 - \sqrt{1+\Delta^2\over 1+2{\cal E}^2+\Delta^2}.
\end{equation}
\end{subequations}
Therefore Eq.~\eqref{alpha} yields different results for the absorbance at circular and linear polarizations:
\begin{equation}
\label{eta_circ_1}
\eta_\text{circ}= {\eta_0\over \sqrt{1+{\cal E}^2\tau_\varepsilon/\tau}},
\end{equation}
\begin{align}
\label{eta_lin_1}
&\eta_\text{lin} = {4\eta_0\over \pi}\\
& 
\times \int\limits_0^\infty  {d\Delta\over 2{\cal E}^2\tau_\varepsilon/\tau + 1 +\Delta^2 + \sqrt{(1+2{\cal E}^2 +\Delta^2)(1 +\Delta^2)} }. \nonumber
\end{align}
This means that the linear-circular dichroism of absorption in 3D topological insulators emerges in the nonlinear in intensity regime.

\begin{figure}[h]
	\centering
		\includegraphics[width=0.9\linewidth]{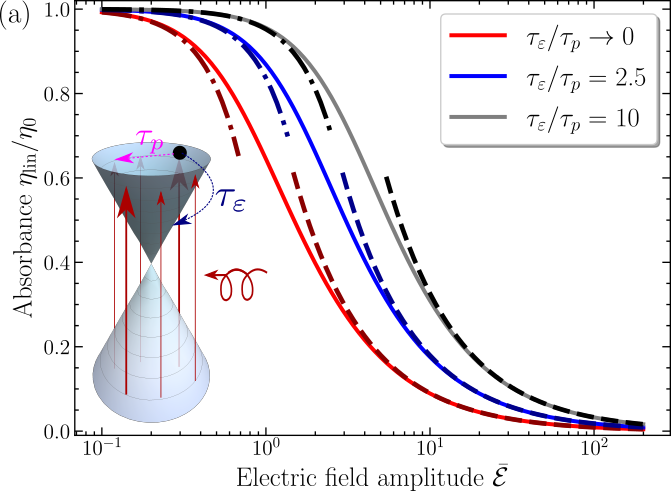} 
		\includegraphics[width=0.9\linewidth]{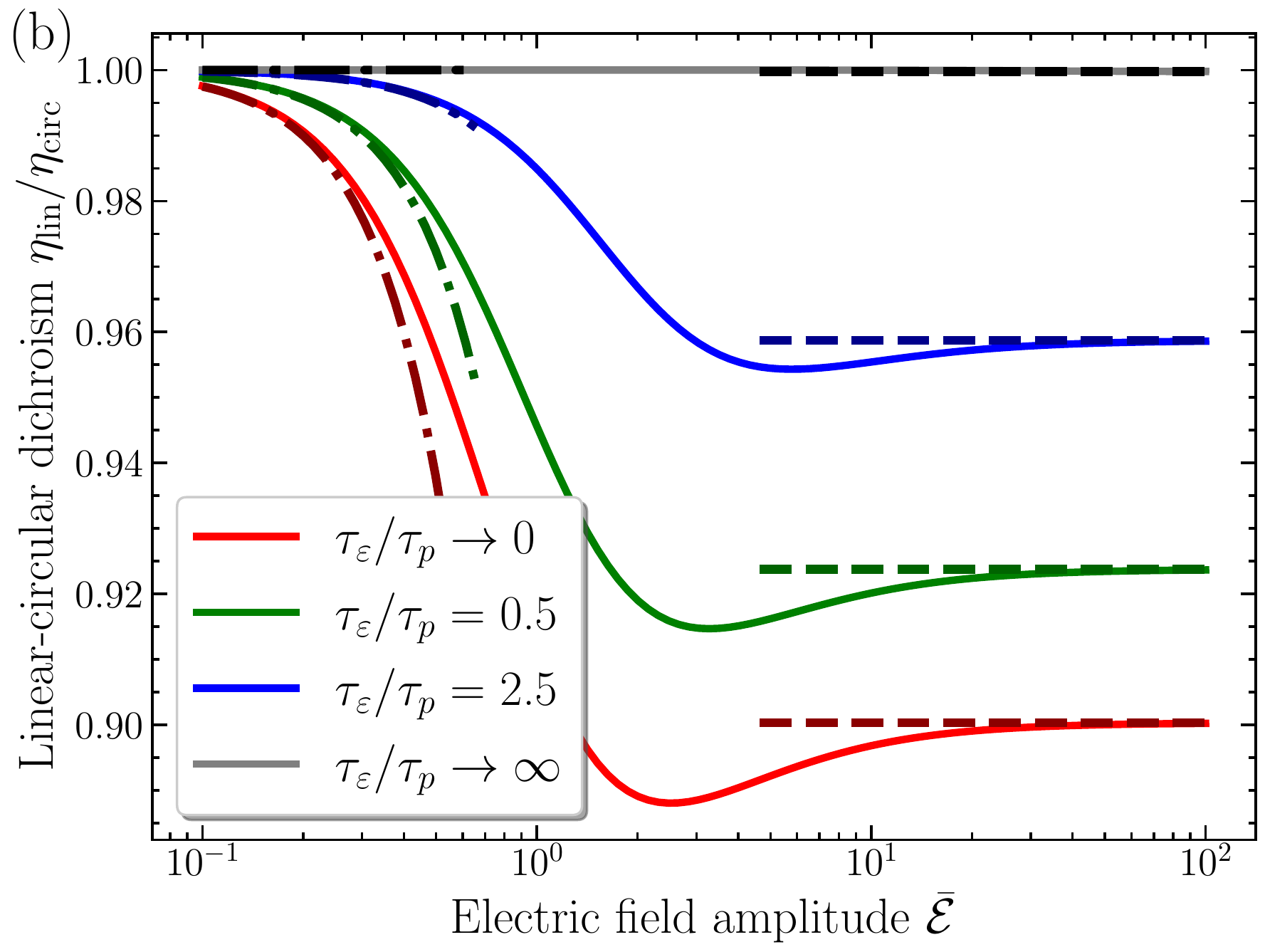}
	\caption{(a): One-photon absorbance at transitions between 2D surface states at linear light polarization.
	Inset shows optical transitions under linearly polarized light and relaxation processes for photoexcited electrons. Dash-dotted and dashed lines are the asymptotes~\eqref{eq:2D_one_phot_small} and~\eqref{high_I_limit_1}, respectively.
	 (b): The linear-circular dichroism degree $\eta_\text{lin}/\eta_\text{circ}$ 
	for various ratios of the relaxation times. Dash-dotted and dashed lines show the asymptotes~\eqref{eq:dich_low_int} and~\eqref{eq:dich_high_int}.}
	\label{fig_LCD}
\end{figure}

To highlight the effect of the elastic scattering we study the dependencies of the absorbance 
%$\eta_\text{lin}$ and $\eta_\text{circ}$ 
on the dimensionless electric field amplitude that is independent of $\tau_p$:
\begin{equation}\label{eq:bar_E_2D}
	\bar{\mathcal{E}} = \frac{\tau_\varepsilon}{\tau}\mathcal{E}.
\end{equation}
The absorbance of circularly-polarized light, Eq.~\eqref{eta_circ_1}, is given by $\eta_\text{circ}/\eta_0=\qty(1+\bar{\mathcal E}^2\tau/\tau_\varepsilon)^{-1/2}$.
In Fig.~\ref{fig_LCD}, the dependencies $\eta_\text{lin}(\bar{\mathcal E})$ are shown for various values of the parameter $\tau_\varepsilon/\tau_p$. 
It demonstrates a strong sensitivity of the absorbance to the relaxation times ratio in the nonlinear in intensity regime.

The low-intensity expansions read as
\begin{subequations}\label{eq:2D_one_phot_small}
\begin{equation}
	\frac{\eta_\text{circ}}{\eta_0} \approx 1-\frac{\bar{\mathcal{E}}^2\tau}{2\tau_\varepsilon}+{\cal O}\qty(\bar{\mathcal{E}}^4),
\end{equation} 
\begin{equation}
	\frac{\eta_\text{lin}}{\eta_0} \approx 1-\frac{\bar{\cal E}^2\tau}{2\tau_\varepsilon}\qty(1+\frac{\tau}{2\tau_\varepsilon})+{\cal O}\qty(\bar{\mathcal{E}}^4).
\end{equation}
\end{subequations}

At high intensity, when $\bar{\cal E} \gg 1, \sqrt{\tau/\tau_\varepsilon}$, the absorbance drops as $1/\sqrt{I}$ at both circular and linear polarizations:
\begin{equation}
\label{high_I_limit_1}
{\eta_\text{circ}\over \eta_0}
\approx {\sqrt{\tau_\varepsilon/\tau}\over \bar{\cal E}},
\qquad
{\eta_\text{lin}\over \eta_0}\approx {\phi(\tau_\varepsilon/\tau_p)\over \bar{\cal E}},
\end{equation}
where
\begin{equation*}
\phi(t) = {2\sqrt{2} (1+t) \over \pi}\qty{ {t\qty[\pi-\arctan\qty({\sqrt{2t+1}\over t})]\over (2t+1)^{3/2}}
+ {1\over 2t+1}}.
\end{equation*}

Figure~\ref{fig_LCD}(b) shows the degree of the linear-circular dichroism, $\eta_\text{lin}/\eta_\text{circ}$.
It differs from unity in the nonlinear in intensity regime by a value $\lesssim 10$~\%. The deviation 
is maximal at some intermediate electric field $\bar{\mathcal E} \gtrsim 1$ which changes with $\tau_\varepsilon/\tau_p$. The low intensity limits~\eqref{eq:2D_one_phot_small} yield the expression for the dichroism 
\begin{equation}\label{eq:dich_low_int}
	\frac{\eta_{\text{lin}}}{\eta_\text{circ}} = 1-\frac{\bar{\cal E}^2\tau^2}{4\tau_\varepsilon^2}+{\cal O}\qty(\bar{\cal E}^4).
\end{equation}
At high intensity, the asymptotes of $\eta_\text{circ,lin}(\bar{\cal E}\to \infty)$ coincide at ${\tau_\varepsilon \gg \tau_p}$: they both are given by $\eta/\eta_0\approx{\sqrt{\tau_\varepsilon/\tau_p}/ \bar{\cal E}}$. Combining this with Eq.~\eqref{eq:dich_low_int} we obtain an absence of the dichroism in the case of fast momentum relaxation.

By contrast, at $\tau_\varepsilon \ll \tau_p$ we obtain $\eta_\text{lin}/\eta_\text{circ} \approx 1-\bar{\cal E}^2/4$  and 
$\eta_\text{lin}/\eta_\text{circ} = 2\sqrt{2}/\pi\approx 0.9$ for the low and high intensities, respectively. Thus, the maximal dichroism is observed in this case. For an arbitrary relation between $\tau_\varepsilon$ and $\tau_p$ the high-intensity dichroism degree is given by 
\begin{equation}\label{eq:dich_high_int}
	\frac{\eta_\text{lin}}{\eta_\text{circ}} = \frac{\phi(\tau_\varepsilon/\tau_p)}{\sqrt{\tau_\varepsilon/\tau}}.
\end{equation}
Figure~\ref{fig_LCD}(b) demonstrates that asymptotics~\eqref{eq:dich_low_int} and~\eqref{eq:dich_high_int} are valid at  $\bar{\mathcal E} \lesssim 0.3$ and $\bar{\mathcal E} \gtrsim 20$, respectively.

\subsection{Two-photon absorption}

One-photon transitions are forbidden while two-photon absorption is allowed in doped samples if the photon energy lies in the range $\varepsilon_{\rm F} < \hbar\omega < 2\varepsilon_{\rm F}$ with $\varepsilon_{\rm F}$ being the Fermi energy.
Two-photon absorption in 2D surface states is very similar to that in graphene where it is well known.
The two-photon absorption matrix element squared is given by~\cite{edge_nonlinear}
\begin{equation}
\label{M_2_square}
\qty|M_{cv}^{(2)}|^2 =   {(ev_0)^4 \over \hbar^2 \omega^6} E_0^4(1-P_\text{lin}^2\cos^2{2\varphi}).
\end{equation}
This yields the low-intensity two-photon absorbance $\eta_0^{(2)}$ in the form~\cite{TPA_BLG,TPA_gr_BLG}
\begin{equation}
\label{eta_0_2}
\eta_0^{(2)} =I \qty({e^2\over \hbar c})^2 {4\pi^2 v_0^2 \over \hbar \omega^4} (1-P_\text{lin}^2/2) .
\end{equation}

Now let us turn to the nonlinear in intensity regime of two-photon absorption. 
It follows from Eq.~\eqref{M_2_square} that the dimensionless electric field amplitude Eq.~\eqref{E} for two-photon absorption, ${\cal E}_2$, for the general case of elliptically polarized radiation is given by
\begin{equation}
{\cal E}_2 = {2\tau(ev_0)^2 \over \hbar^2 \omega^3} E_0^2 \sqrt{1-P_\text{lin}^2/2}.
\end{equation}
This means that ${\cal E}_2$ is different for circularly and linearly polarized radiation:
\begin{equation}
\label{E_2_circ_lin}
{\cal E}_{2,\text{circ}} = {2\tau(ev_0)^2 \over \hbar^2 \omega^3} E_0^2,
\qquad
{\cal E}_{2,\text{lin}} = {{\cal E}_{2,\text{circ}} \over \sqrt{2}}.
\end{equation}
Then from Eq.~\eqref{Psi} we obtain
\begin{subequations}
\label{Psi_2}
\begin{equation}
\label{Psi_2_circ}
\Psi_\text{circ}^{(2)} = {{\cal E}_{2,\text{circ}}^2  \over 1+{\cal E}_{2,\text{circ}}^2+\Delta^2},
\end{equation}
\begin{equation}
\label{Psi_2_lin}
\Psi_\text{lin}^{(2)} =1 - \sqrt{1+\Delta^2\over 1+2{\cal E}_{2,\text{lin}}^2+\Delta^2}.
\end{equation}
\end{subequations}
The expressions~\eqref{Psi_2} pass into Eqs.~\eqref{Psi_1} after substitutions ${\cal E}_{2,\text{circ}} \to {\cal E}$ and  ${\cal E}_{2,\text{lin}} \to {\cal E}$, respectively.
Therefore the intensity dependencies of $\eta_{\text{circ}}^{(2)}/\eta_{0,\text{circ}}^{(2)}$ and $\eta_{\text{lin}}^{(2)}/\eta_{0,\text{lin}}^{(2)}$ are given by Eqs.~\eqref{eta_circ_1} and~\eqref{eta_lin_1} with the same substitutions. The high-intensity limits follow from Eqs.~\eqref{high_I_limit_1}:
\begin{equation}
\label{high_I_value_2ph}
\eta_\text{circ}^{(2)} \approx {\pi e^2\over\hbar c}{1\over \omega \sqrt{\tau\tau_\varepsilon}},
\qquad
\eta_\text{lin}^{(2)} \approx {\pi e^2\over  \hbar c}{\phi(\tau_\varepsilon/\tau_p)\over {\sqrt{2}} \omega \tau_\varepsilon}.
\end{equation}
%\begin{subequations}\label{high_I_value_2ph}
%\begin{equation}
%\eta_\text{circ}^{(2)}(I\to \infty)
% = {\pi e^2\over\hbar c}{1\over \omega \sqrt{\tau\tau_\varepsilon}},
%\end{equation}
%\begin{equation}
%\eta_\text{lin}^{(2)}(I\to \infty)
%= {\pi e^2\over  \hbar c}{\phi(\tau_\varepsilon/\tau_p)\over {\sqrt{2}} \omega \tau}.
%\end{equation}
%\end{subequations}
This means that the two-photon absorbance tends to a constant at 
%high intensity 
$I\to \infty$
at both circular and linear polarizations. 

\begin{figure}[b]
	\centering
	\includegraphics[width=0.9\linewidth]{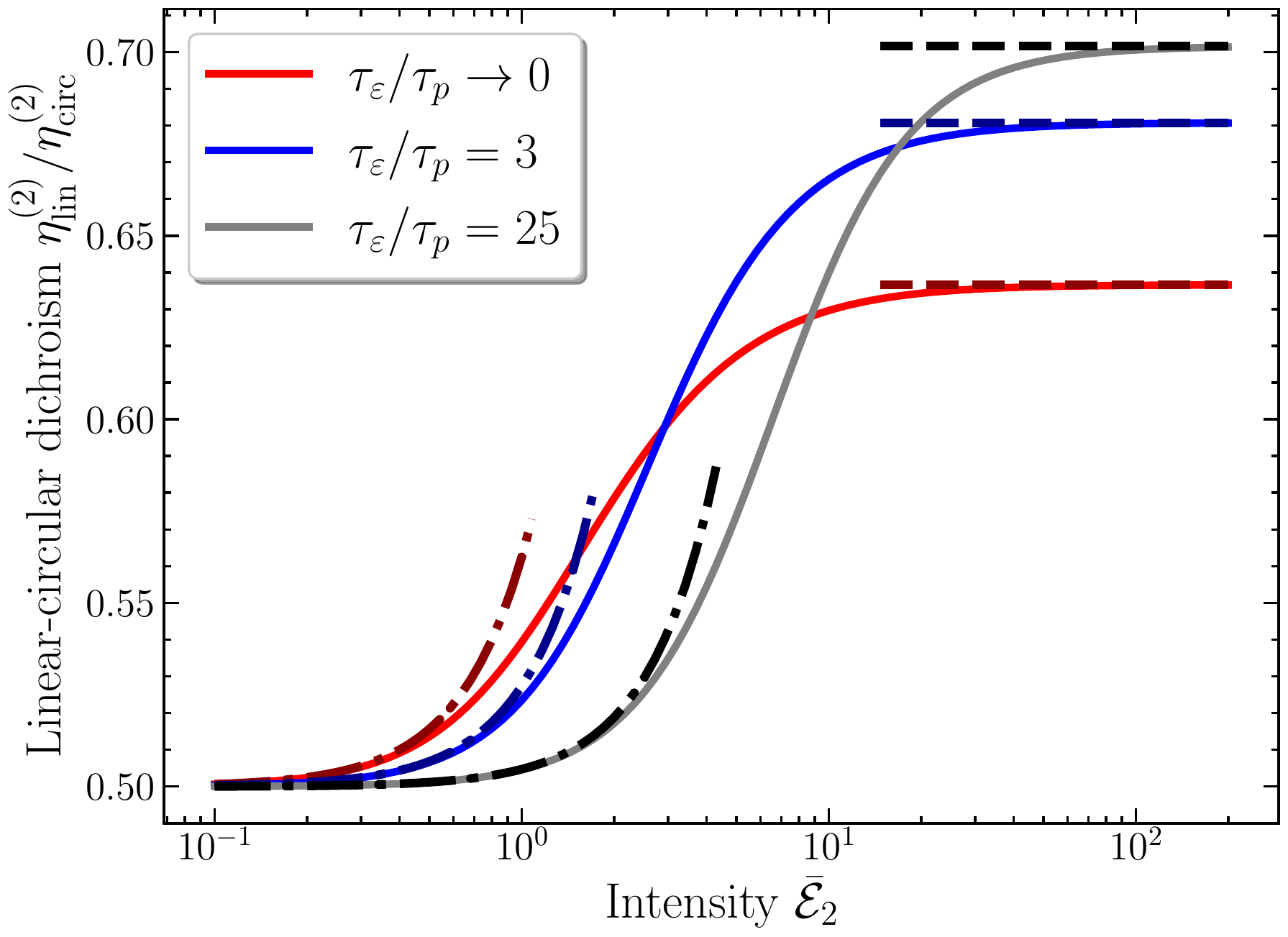} 
	\caption{Linear-circular dichroism of two-photon absorption
	%~$\eta_{\text{lin}}^{(2)}/\eta_{\text{circ}}^{(2)}$ as a function of the dimensionless light intensity
	%~$\bar{\mathcal{E}}_{2}$ 
	for various ratios of the relaxation times. The saturation values obtained from Eq.~\eqref{high_I_value_2ph} are shown by dashed lines. Dash-dotted lines show the low-intensity expansion~\eqref{eq:ans}.}
	\label{fig:2_phot_dich}
\end{figure}

The intensity dependence of the linear-circular dichroism in two-photon absorption is presented in Fig.~\ref{fig:2_phot_dich}. 
We again
introduce the dimensionless light intensity 
$$\bar{\mathcal{E}}_{2} = \mathcal{E}_{2,\text{circ}}\tau_\varepsilon/\tau$$ that is independent of $\tau_p$.
Figure~\ref{fig:2_phot_dich} demonstrates that the dichroism degree increases with intensity from the value $1/2$ following from Eq.~\eqref{eta_0_2} to a 
%higher \NL{lower?} 
closer to unity value which is governed by the relaxation-time ratio. The high-intensity limit is realized at $\bar{\mathcal E}_{2} \approx 10$. 
It follows from the asymptotes~\eqref{high_I_value_2ph} that the absorbance ratio at $\tau_p\ll \tau_\varepsilon$ is equal to $1/\sqrt{2} \approx 0.71$, while in the opposite limit $\tau_p \gg \tau_\varepsilon$ the saturation value $2/\pi\approx 0.64$.

At low intensity, the first correction to the degree of the linear-circular dichroism reads
\begin{equation}\label{eq:ans}
	\frac{\eta_{\text{lin}}^{(2)}}{\eta_{\text{circ}}^{(2)}}  \approx 
%	\frac{\eta_{\text{0,lin}}^{(2)}}{\eta_{\text{0,circ}}^{(2)}}\qty[1+\bar{\mathcal{E}}_2^2\frac{\tau}{4\tau_\varepsilon}\qty(1-\frac{\tau}{2\tau_\varepsilon})] = 
	\frac{1}{2}\qty[1+\bar{\mathcal{E}}_2^2\frac{\tau}{4\tau_\varepsilon}\qty(1-\frac{\tau}{2\tau_\varepsilon})].
\end{equation}
In the limit $\tau_\varepsilon/\tau_p\to 0$ this yields the quadratic in intensity dependence $\eta_{\text{lin}}^{(2)}/\eta_{\text{circ}}^{(2)}\approx \frac{1}{2}\qty(1+{\bar{\mathcal{E}}_2^2/8})$. By contrast, at efficient elastic scattering when $\tau_\varepsilon/\tau_p\to \infty$, the nonlinear regime appears in the next order in intensity only: $\eta_{\text{lin}}^{(2)}/\eta_{\text{circ}}^{(2)}= 1/2 + \mathcal O(I^4)$.

\section{1D edge states}
\label{1DES}

%\NL{\sout{Hamiltonian?}}

1D states at edges of 2D topological insulators  are characterized by a 1D wavevector $k$ along the edge.
Electron and hole states could be obtained from  Bernevig-Hughes-Zhang Hamiltonian~\cite{TI_book}. 
Energy spectrum of 1D edge states $\varepsilon_{c,v}=\pm \hbar v_0 k$ is shown in Fig.~\ref{fig:sketch}(a).
Hereafter, the electron-hole symmetry is assumed.
%\commentNL{More links on the edge states obtained from the same bulk Hamiltonian?}\LG{Yes!}

The absorbed power $wI/(\hbar\omega)$ is described in 1D systems by the absorption length $w$. The absorption nonlinearity is obtained from Eq.~\eqref{alpha} with the function $\Psi(\Delta)$ given by Eq.~\eqref{Psi} where instead of angular averaging one performs summation over two signs of the wavevector:
\begin{equation}
\left< \ldots \right> \to {1\over 2}\sum_{k=\pm \abs{k}} \ldots \nonumber
\end{equation}
Due to the electron-hole symmetry, the equilibrium occupation difference $\mathcal{F}$ is given by Eq.~\eqref{F}
and relaxation times in the conduction and valence bands are equal, so the total relaxation rate $1/\tau$ is given by Eq.~\eqref{inverse_tau}.

The symmetry of the system allows for generation of photocurrents at the moment of excitation~\cite{DurnevAnnPhys}.
The photocurrent density is given by
\begin{equation}
j  = e v_0 \sum_{k} \text{sgn}(k) (f_c-f_v).
\end{equation}
Here we took into account that the electron velocity in the conduction band $v_c(k)=v_0 \text{sgn}(k)$, and in the valence band $v_v(k)=-v_c(k)$. 
Then, from the expression for nonequilibrium occupations~\eqref{fcv1} we obtain:
\begin{equation}
j  
= 2ev_0 \mathcal{F} \sum_{k>0} {\Phi  \over 1+\Psi  \tau_\varepsilon/\tau_p}.
\end{equation}
Here we introduce the function describing a difference in the generation rates at $k=|k|$ and $k=-|k|$:  
\begin{equation}
\label{Phi_def}
\Phi  
= {1\over 2}\sum_{k=\pm|k|} {\text{sgn}(k)G \over 1/2\tau + G }
.
\end{equation}
%
%Making the same assumptions as at derivation of the absorption length, we get
Similarly to the derivation of the absorption length, we get
\begin{equation}
\label{j_Delta_int}
j=  {e{\cal F} \over  \pi \tau}\int\limits_0^\infty d\Delta {\Phi \over 1+\Psi \tau_\varepsilon/\tau_p}.
\end{equation}

Optical absorption in the Bernevig-Hughes-Zhang model is very weak since it is possible via a magneto-dipole mechanism only~\cite{Dora2012,Artemenko2013,ArtemenkoPRB,Junck2013}.
The situation changes if the symmetry-enforced level anticrossing  at interfaces is taken into account~\cite{Zunger_gamma}. It results in a  
%It was shown~\cite{Zunger_gamma}, that symmetry-enforced level anticrossing  at interfaces modifies the energy spectrum. Such modification is the reason of the 
mixing of bulk electron and holes subbands 
%and breaking of the inversion symmetry, 
that 
%leads to an appearance of 
makes possible
electro-dipole transitions between the edge states~\cite{DurnevAnnPhys,Durnev_JPCM}.
Below we consider absorption and photocurrent first at circular polarization and then for linearly-polarized light.

\subsection{Circular polarization}

At circular polarization, the matrix element for direct optical transitions between the edge states in 2D topological insulators is nonzero in the electro-dipole approximation. It is given by~\cite{Durnev_JPCM}
%\begin{equation}
%\label{M_quad_circ}
%|M|^2_\text{circ} = E_0^2 \qty[{D_1^2+D_2^2\over 2} - \text{sign}(k)D_1D_2 P_\text{circ}o_z]k^2,
%\end{equation}
\begin{equation}
\label{M_quad_circ}
|M|^2_\text{circ} = E_0^2 {D_1^2+D_2^2\over 2}\qty[1 + \text{sgn}(k) \xi P_\text{circ}o_z]k^2,
\end{equation}
where $\bm o$ is the unit vector in the light propagation direction, $z$ is the direction of the normal to the sample, $P_\text{circ}=i(\bm E_0 \times \bm E_0^*)_z o_z/E_0^2$ is the 
%circular polarization degree
light helicity,  
%\LG{$o_z$?}
%\commentNL{
%%I replaced $\bm e$ in $P_\text{circ}$. 
%I think we could left $o_z$ to derive formulas in the most general case.} 
and $D_{1,2}$ are two linearly-independent constants giving rise to the $k$-linear dipole momentum matrix elements. 
The asymmetry factor reads as
\begin{equation}
\xi =-{2D_1D_2\over D_1^2+D_2^2}.
\end{equation}

\subsubsection{Absorption}

The low-intensity absorption length 
is given by~\cite{Durnev_JPCM}
\begin{equation}
w_{0,\text{circ}} = {\cal F}(D_1^2+D_2^2) \qty({\omega\over v_0})^3{\pi\over4\hbar c} \equiv \overline{w}_0.
\end{equation}
%\begin{multline}
%w_{0,\text{circ}} = {\cal F}(D_1^2+D_2^2) \qty({\omega\over v_0})^3{\pi\over4\hbar c n_\omega} \equiv \overline{w}_0,
%\\
%j_{0,\text{circ}} = -4e v_0\tau {D_1D_2\over D_1^2+D_2^2} P_\text{circ}o_z {\overline{w}_0 I\over \hbar\omega}\equiv \overline{j}_0.
%\end{multline}

It follows from Eqs.~\eqref{M_quad_circ} and~\eqref{Psi} that
\begin{equation}
\label{Psi_w}
\Psi_\text{circ}  
= {1+(1+\Delta^2)/{\cal E}_\text{1D}^2-\xi^2\over [1+(1+\Delta^2)/{\cal E}_\text{1D}^2]^2-\xi^2},
\end{equation}
%\begin{equation}
%\label{Psi_w}
%\Psi  
%= {A+(\Delta/{\cal E})^2-\xi^2\over [A+(\Delta/{\cal E})^2]^2-\xi^2},
%\end{equation}
where $\mathcal E_\text{1D}$ is introduced according to Eq.~\eqref{E}:
\begin{equation}
\label{E_circ_1D}
{\cal E}_\text{1D}={E_0 {\omega}\tau  \sqrt{D_1^2+D_2^2}\over {\sqrt{2} v_0} \hbar}.
%, \quad \Xi=\qty({2D_1D_2\over D_1^2+D_2^2})^2.
\end{equation}

Therefore we have
from Eq.~\eqref{alpha} the absorption length in the form
\begin{equation}
\label{w_w0}
{w_\text{circ}\over \overline{w}_0}={1\over 2R}\sum_\pm {R \pm (\xi^2+t/2) \over \sqrt{1+{\cal E}_\text{1D}^2(1+t/2 \pm R)}}.
\end{equation}
Here 
\begin{equation}
\label{tR}
t = {\tau_\varepsilon \over \tau_p}, \qquad R=\sqrt{t^2/4+\xi^2(1+t)}.
\end{equation}

At large light intensities we have 
$w/\overline{w}_0 \sim 1/{\cal E}_\text{1D} \sim 1/\sqrt{I}$.
At $\tau_\varepsilon/\tau_p \to \infty$, 
we obtain for all intensities
\begin{equation}\label{w_w0_appr}
{w_\text{circ}(\tau_\varepsilon/\tau_p \to \infty) \over \overline{w}_0} =  {1\over {\cal E}_\text{1D}}\sqrt{\tau_p \over \tau_\varepsilon}.
\end{equation}

\begin{figure}[t]
\centering
\includegraphics[width=0.9\linewidth]{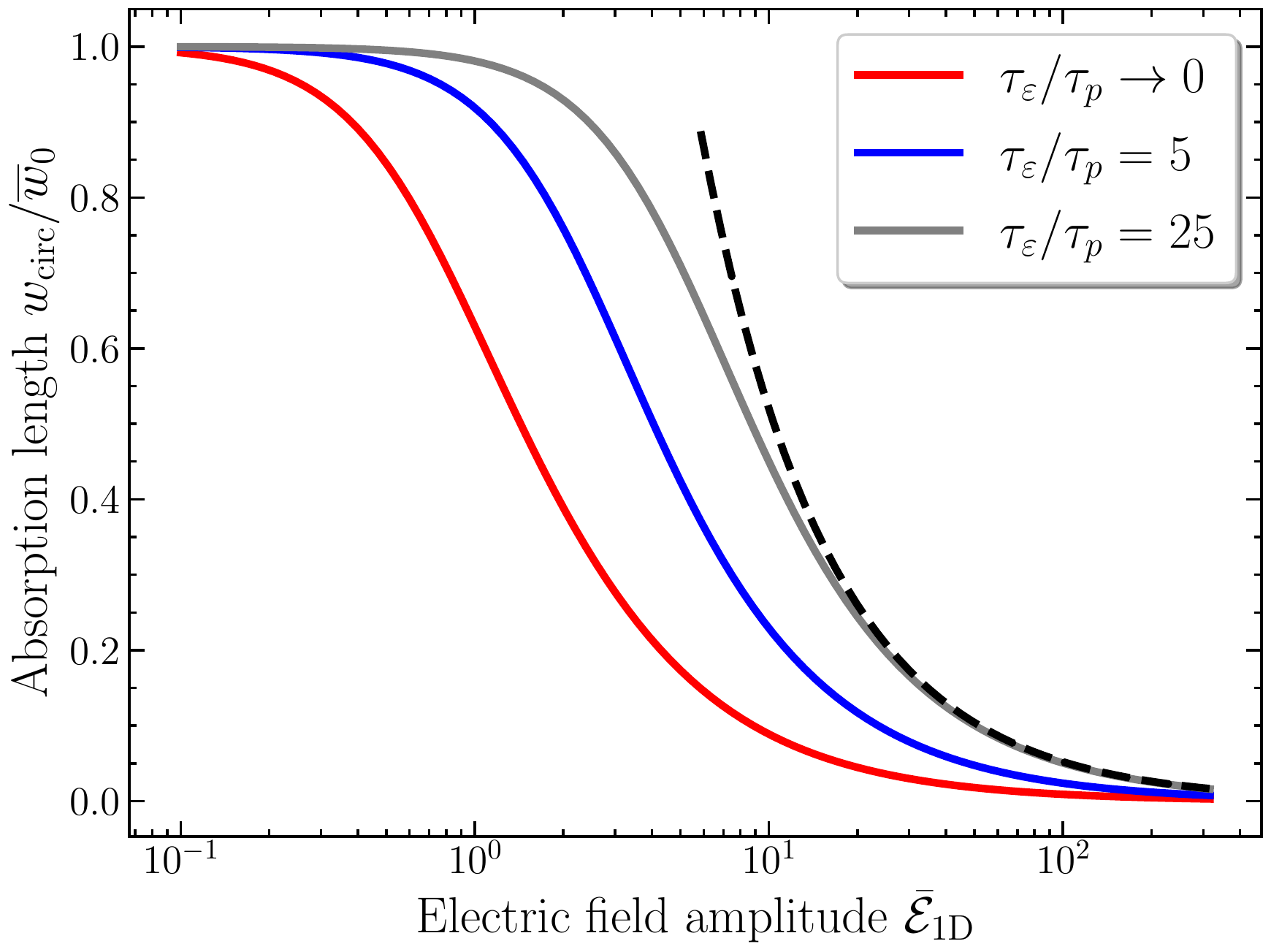} 
\caption{Absorption length at circular polarization for various ratios of the energy and momentum relaxation times.
%$w_{\text{circ}}/\overline{w}_0$  
Calculations are performed for
the parameters of edge states in HgTe/CdHgTe quantum wells.
The dashed line shows the asymptotics~\eqref{w_w0_appr} for ${\tau_\varepsilon/\tau_p =  25}$.}
\label{fig:abs_circ}
\end{figure}

Similar to the case of the 2D surface states, 
we introduce the dimensionless electric field amplitude that is a product of the Rabi frequency and energy relaxation time 
\begin{equation}
	\bar{\mathcal{E}}_{\text{1D}} = \frac{\tau_\varepsilon}{\tau} \mathcal{E}_{\text{1D}}.
\end{equation}
In Fig.~\ref{fig:abs_circ} we plot the dependence of the absorption length on $\bar{\mathcal{E}}_{\text{1D}}$ 
for various ratios of the energy and momentum relaxation times.
%The absorption length dependencies on the electric field amplitude 
%for various ratios of the energy and momentum relaxation times are shown in Fig.~\ref{fig:abs_circ}.
In calculations in this Section we  take $D_1/D_2= 0.5$, $\abs{\xi}=0.8$ corresponding to 1D edge states in HgTe/CdHgTe quantum wells~\cite{Durnev_JPCM}. It is seen from Fig.~\ref{fig:abs_circ} that the nonlinear regime of absorption emerges at $\bar{\mathcal E}_\text{1D} \gtrsim 0.1$. The dependence of $w_\text{circ}$ on the electric field amplitude follows the asymptotics~\eqref{w_w0_appr} at $\bar{\mathcal E}_\text{1D} \gtrsim 10$.

\subsubsection{Photocurrent}

At circular polarization, the photocurrent is generated at light absorption in 1D edge states which is caused by the Circular Photogalvanic Effect (CPGE).
For its calculation 
%of the circular photocurrent 
we use Eq.~\eqref{j_Delta_int}
where $\Psi$ is given by Eq.~\eqref{Psi_w}, and
\begin{equation}
\Phi_\text{circ}= P_\text{circ}o_z \xi {1+(1+\Delta^2)/{\cal E}_\text{1D}^2-1\over [1+(1+\Delta^2)/{\cal E}_\text{1D}^2]^2-\xi^2}.
\end{equation}
As a result, we obtain from Eq.~\eqref{j_Delta_int} 
%the photocurrent 
\begin{equation}
\label{j_j0}
{j_\text{circ}\over j_1}={1\over 2R}\sum_\pm {R  \pm (1+t/2) \over \sqrt{1+{\cal E}_\text{1D}^2(1+t/2\pm R)}}.
\end{equation}
Here $t$ and $R$ are defined by Eqs.~\eqref{tR},
and 
$j_1$ is the linear in the light intensity CPGE current~\cite{Durnev_JPCM,DurnevAnnPhys}  
\begin{equation}
\label{j00}
j_1  
%= {2ev_0\tau  \xi } {\overline{w}_0 I\over \hbar \omega}P_\text{circ}o_z
= {e \mathcal{F}\mathcal E_\text{1D}^2\over 2\tau}\xi P_\text{circ}o_z.
\end{equation}

At high light intensities we have 
$j_\text{circ}/j_1 \sim 1/{\cal E}_\text{1D} \sim 1/\sqrt{I}$, i.e. $j \sim \sqrt{I}$.
At $\tau_\varepsilon/\tau_p \to \infty$, 
we obtain the same asymptotics as for the absorption length:
%$w_\text{circ}/\overline{w}_0$:
\begin{equation}\label{j_j0_approx}
{j_\text{circ}(\tau_\varepsilon/\tau_p \to \infty) \over j_1} =  {1\over {\cal E}_\text{1D}}\sqrt{\tau_p \over \tau_\varepsilon}.
\end{equation}

\begin{figure}[t]
\centering
\includegraphics[width=0.9\linewidth]{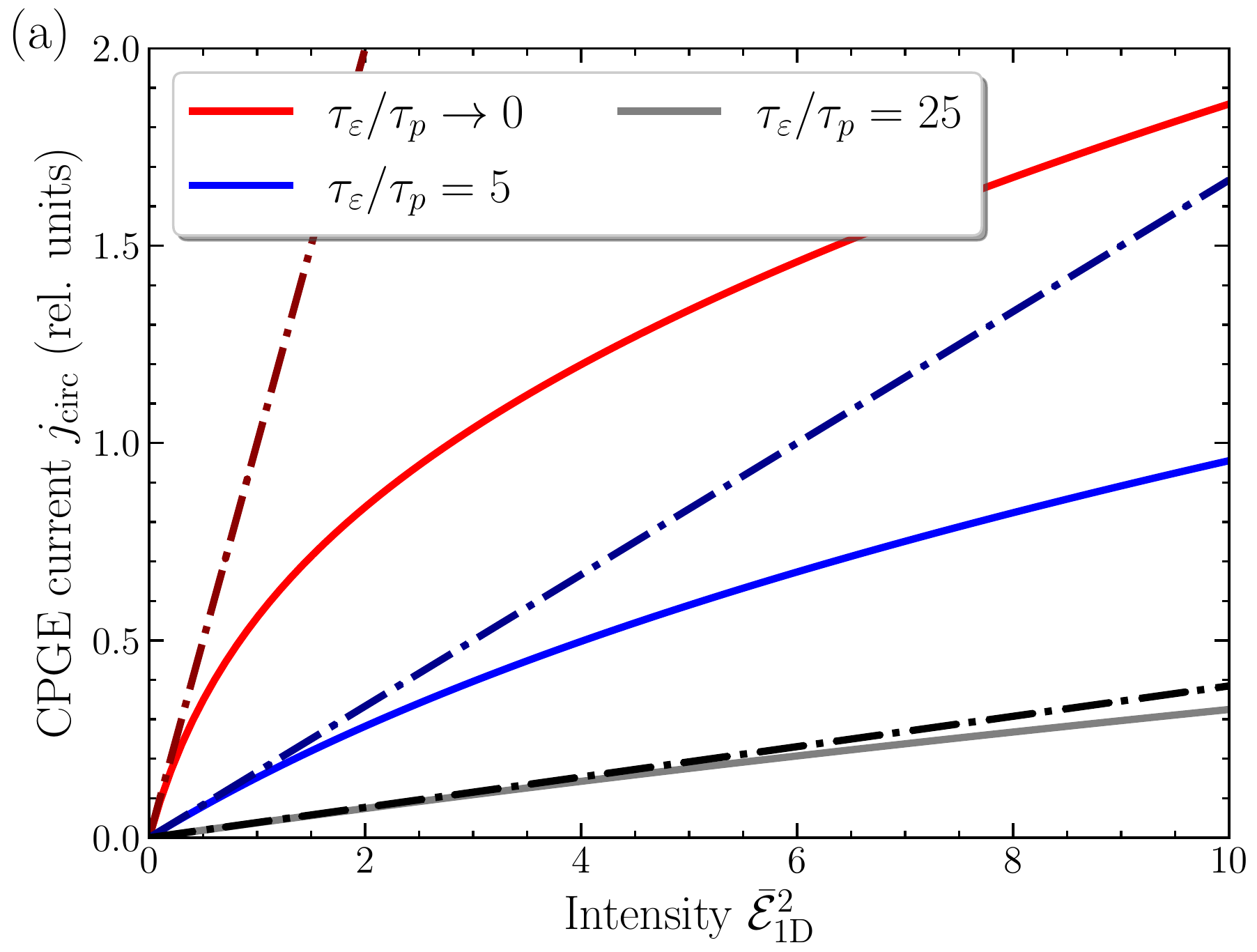} 
\includegraphics[width=0.9\linewidth]{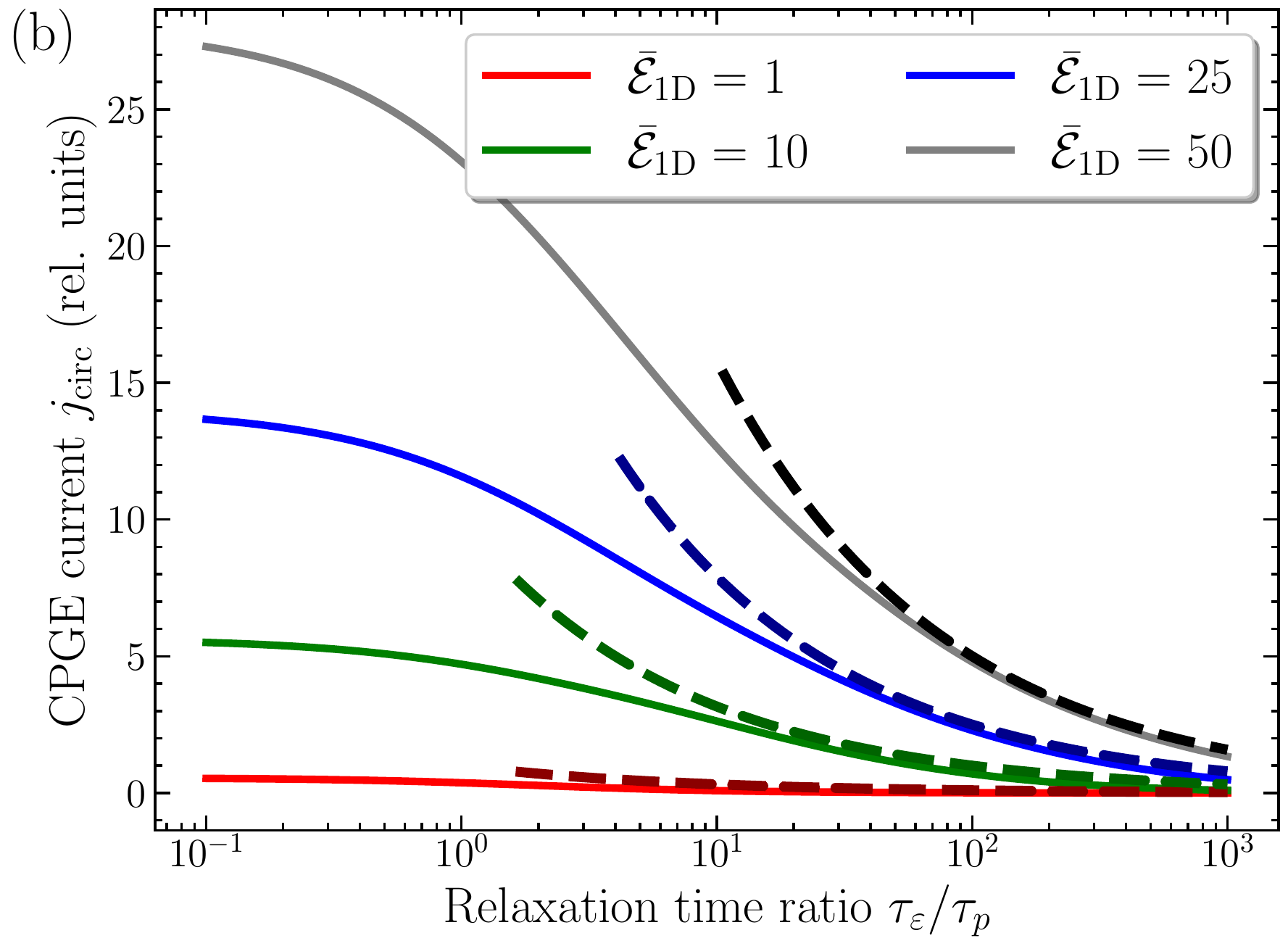} 
\caption{(a) CPGE current dependence on the dimensionless light intensity $\bar{\mathcal{E}}_\text{1D}^2$ at various ratios of the energy and momentum relaxation times. 
The photocurrent is given in the units $e \mathcal{F}\xi/(2\tau_\varepsilon)$.
The dashed-dotted lines shows the linear in the intensity current~\eqref{j00} with the corresponding relaxation time.
%with $\xi = \xi_{\text{circ}}$ \eqref{eq:xi_circ}. 
(b)~Circular photocurrent dependence on the relaxation time ratio at various field amplitudes. The dashed lines show the approximation~\eqref{j_j0_approx}.}
\label{fig:cpge}
\end{figure}

The dependence of the CPGE current $j_\text{circ}$ 
%(in the units $e \mathcal{F}\xi/(2\tau_\varepsilon) = j_1 \tau/(\mathcal{E}_{\text{1D}}^2\tau_\varepsilon)$) 
on the light intensity
%calculated by Eq.~\eqref{j_j0} for 
is shown in Fig.~\ref{fig:cpge}(a) for various $\tau_\varepsilon/\tau_p$. 
%\NL{\sout{The difference of the current from the linear in intensity value 
%$j_1$ is clearly seen starting from $\mathcal{E}_{\text{1D}}^2\sim 0.1$. }}
It demonstrates that the effect of elastic scattering on the circular photocurrent is opposite to that on absorption. It is seen from Fig.~\ref{fig:cpge}(b) that the stronger is momentum scattering the smaller is photocurrent. The asymptotics~\eqref{j_j0_approx} is realized starting from some values of $\tau_\varepsilon/\tau_p$ depending on intensity.

\subsection{Linear polarization}

For direct optical transitions at linear polarization, the  matrix element squared reads as~\cite{Durnev_JPCM}
\begin{align}
&|M|^2_\text{lin} = E_0^2 {D_1^2+D_2^2\over 2}\biggl\{\qty(1 -\zeta\cos{2\alpha})k^2 
- \mu_\text{B} o_zk \nonumber
\\
&\times \qty[{D_1g_2-D_2g_1\over D_1^2+D_2^2}\cos{2\theta}+
{D_1g_2+D_2g_1\over D_1^2+D_2^2}\cos{2(\alpha-\theta)}]
\biggr\}. 
\end{align}
Here $\alpha$ is an angle between the light polarization direction and the normal to the edge in the sample plane, see inset to Fig.~\ref{fig:lin_dich}, and $\theta$ is an angle between the edge and the $[100]$ axis provided the sample is grown along the $z \parallel [001]$ direction.
%, $\bm o$ is the unit vector in the light propagation direction. The constants $D_{1,2}$ and $g_{1,2}$ describe the electro-dipole and magneto-dipole electron-photon interaction mechanisms, respectively.
The second, asymmetrical term, is due to an interference of the electro-dipole and magneto-dipole interactions. 
%Therefore the asymmetry parameter $\xi_\text{lin} \propto 1/c$ is small.
We also introduced the value
\begin{equation}
\zeta = {D_2^2-D_1^2\over D_2^2+D_1^2}.
\end{equation}

\subsubsection{Dichroic absorption}

The linear absorption dichroism, i.e. a dependence of the absorption width on the linear polarization state, is present already in the linear in intensity regime~\cite{Durnev_JPCM}:
\begin{equation}\label{lin_dich}
w_{0,\text{lin}} = \overline{w}_0 \qty(1-\zeta\cos{2\alpha}).
\end{equation}

For calculation of nonlinear absorption, it is enough to use the electro-dipole approximation where we have
% $\xi=0$ and
\begin{equation}
{\cal E}_\text{lin}(\alpha)
={\cal E}_\text{1D}\sqrt{1-\zeta\cos{2\alpha}}
\end{equation}
with $\mathcal E_\text{1D}$ is given by Eq.~\eqref{E_circ_1D},
and
\begin{equation}
\label{Psi_lin}
\Psi_\text{lin}={1\over 1+ (1+\Delta^2)/\mathcal E_\text{lin}^2.}
\end{equation}
The absorption nonlinearity is then obtained from Eq.~\eqref{alpha}:
\begin{equation}
\label{abs_lin_pol}
{w_\text{lin}\over w_{0,\text{lin}}}={1\over \sqrt{1+{\cal E}_\text{lin}^2 \tau_\varepsilon/\tau}}.
\end{equation}

Similar to the case of the circular polarization at large light intensities we have 
$w_{\text{lin}}/w_{0,\text{lin}} \sim 1/{\cal E}_\text{lin} \sim 1/\sqrt{I}$.

\begin{figure}[t]
	\includegraphics[width=0.9\linewidth]{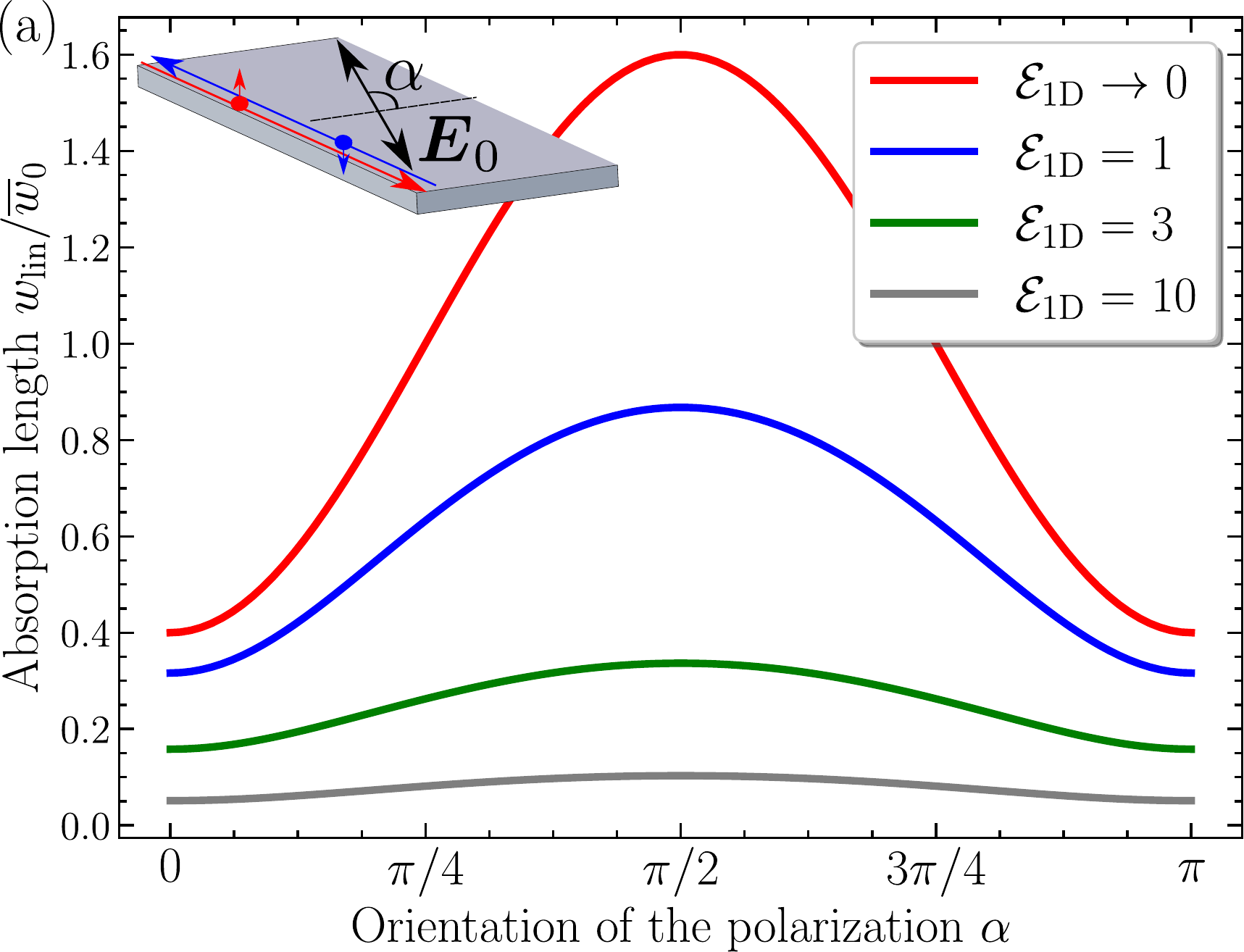} 
	\includegraphics[width=0.9\linewidth]{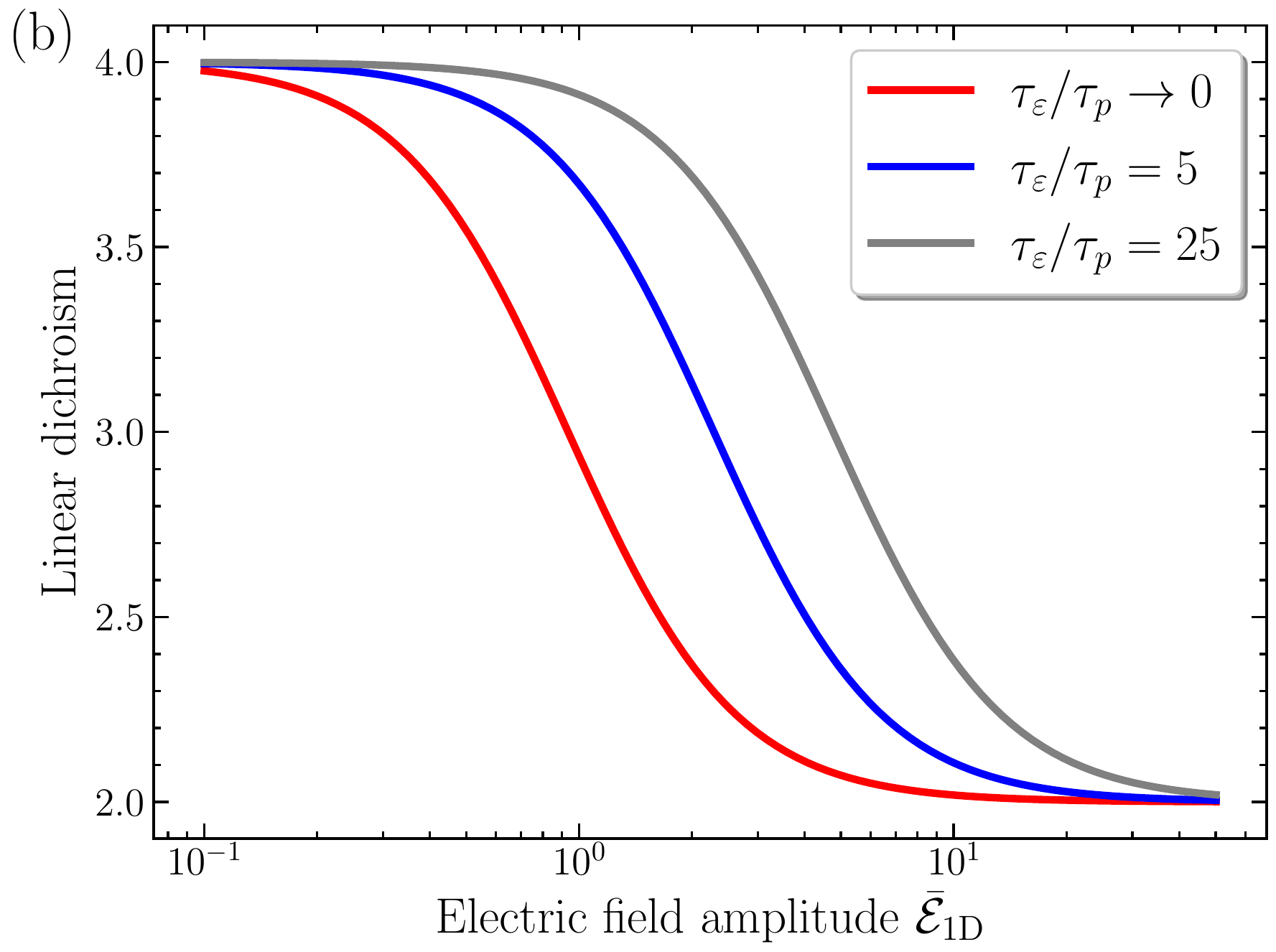} 
	\caption{(a) Angular dependence of the absorption length calculated after Eq.~\eqref{eq:lin_alp} at $\tau_\varepsilon/\tau_p = 0.5$. Inset: Edge of the 2D topological insulator and orientation of the polarization. (b)~Dependence of the linear dichroism~\eqref{eq:lin_dich}  on the dimensionless electric field amplitude %\NL{$\bar{\mathcal{E}}_{\text{1D}}$} 
	at various ratios of the relaxation times. }
	\label{fig:lin_dich}
\end{figure}

Figure~\ref{fig:lin_dich}(a) demonstrates the dependence of the absorption length on the electric field orientation relative to the edge. We normalize the absorption length to $\overline{w}_0$:
\begin{equation}\label{eq:lin_alp}
	\frac{w_{\text{lin}}}{\overline{w}_0} = \frac{1-\zeta\cos{2\alpha}}{\sqrt{1+{\cal E}_{\text{1D}}^2(1-\zeta\cos{2\alpha})\tau_\varepsilon/\tau}}
\end{equation}
and show the dependencies of this ratio on $\alpha$ at various $\mathcal{E}_{\text{1D}}$. 
The largest absorption length is seen to be realized for  all intensities at $\alpha=\pi/2$ 
%(note, that $\zeta<0$)
but its maximal value decreases substantially  with the electric field increase.

It follows from Eqs.~\eqref{lin_dich},~\eqref{abs_lin_pol} that the absorption length' ratio for the light polarization along the edge ($\alpha = \pi/2$) and perpendicular to the edge ($\alpha = 0$) is given by 
\begin{equation}\label{eq:lin_dich}
\frac{w_{\text{lin}}(\alpha = \pi/2)}{w_{\text{lin}}(\alpha = 0)} = \frac{D_2^2}{D_1^2} \sqrt{\frac{1+{\cal E}_{\text{1D}}^2(1-\zeta)\tau_\varepsilon/\tau}{1+{\cal E}_{\text{1D}}^2(1+\zeta)\tau_\varepsilon/\tau}}.
\end{equation}
The dependencies of this ratio on the electric field amplitude for various $\tau_\varepsilon/\tau_p$ are shown in Fig.~\ref{fig:lin_dich}(b). At 
%\NL{$\bar{\mathcal E}_{\text{1D}} \gg 1$}, 
high intensity
this value is independent of the relaxation times ratio being equal to $D_2/D_1$.

\subsubsection{Photocurrent}

The low-intensity photocurrent at linear polarization of light is given by~\cite{Durnev_JPCM,DurnevAnnPhys}
\begin{align}
\label{j_lin1}
&j_{1,\text{lin}} ={j}_1\frac{v_0 \mu_B}{\omega}\\
&\times \frac{(D_1g_2+D_2g_1)\cos 2(\theta-\alpha)+(D_1g_2-D_2g_1)\cos2\theta}{D_1D_2}, \nonumber
\end{align}
where the linear in intensity value $j_1$ is given by Eq.~\eqref{j00}.
This photocurrent is nonzero due to the magneto-dipole interaction. Therefore it belongs to the class of  Photon Drag Effects~\cite{Ivchenko_book}. Since it is generated in the edge of the topological insulator at normal light incidence, it is the transversal photon drag current.

Taking into account the magneto-dipole interaction in the first order, we obtain a nonzero value in the low-intensity regime $j_{1, \text{lin}} \neq 0$. However, the nonlinearity of the photocurrent $j_{\text{lin}}$ is present at linear polarization even in the electro-dipole approximation. 
Therefore we calculate the intensity dependence of $j_{\text{lin}}$ by Eq.~\eqref{j_Delta_int} with 
$\Psi$ from Eq.~\eqref{Psi_lin}. It yields
\begin{equation}
\label{j_lin_pol}
	\frac{j_{\text{lin}}}{j_{1,\text{lin}}} =\frac{\tau_p/\tau}{\sqrt{1+\mathcal{E}_{\text{lin}}^2\tau_\varepsilon/\tau}}-\frac{\tau_p/\tau_\varepsilon}{\sqrt{1+\mathcal{E}_{\text{lin}}^2}}.
\end{equation}
Similarly to the case of the circular photocurrent the high light intensity behavior is $j_{\text{lin}}\sim \sqrt{I}$, and at $\tau_\varepsilon/\tau_p \to \infty$ the asymptotics is 
\begin{equation}
	\frac{j_{\text{lin}}}{j_{1,\text{lin}}} =\frac{1}{\mathcal{E}_{\text{lin}}}\sqrt{\tau_p\over \tau_\varepsilon}.
\end{equation}

\begin{figure}[h]
	\includegraphics[width=0.9\linewidth]{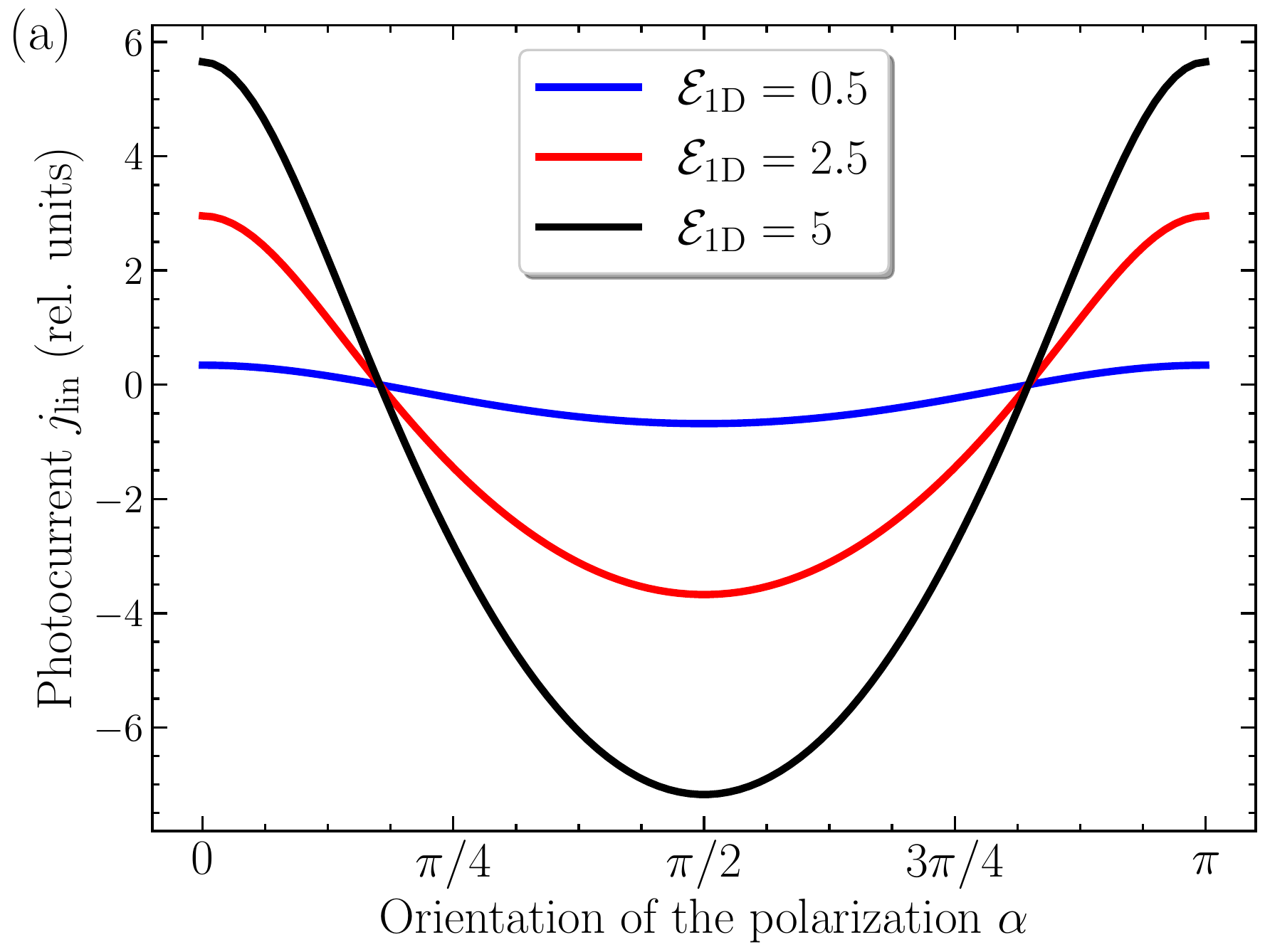} 
	\includegraphics[width=0.9\linewidth]{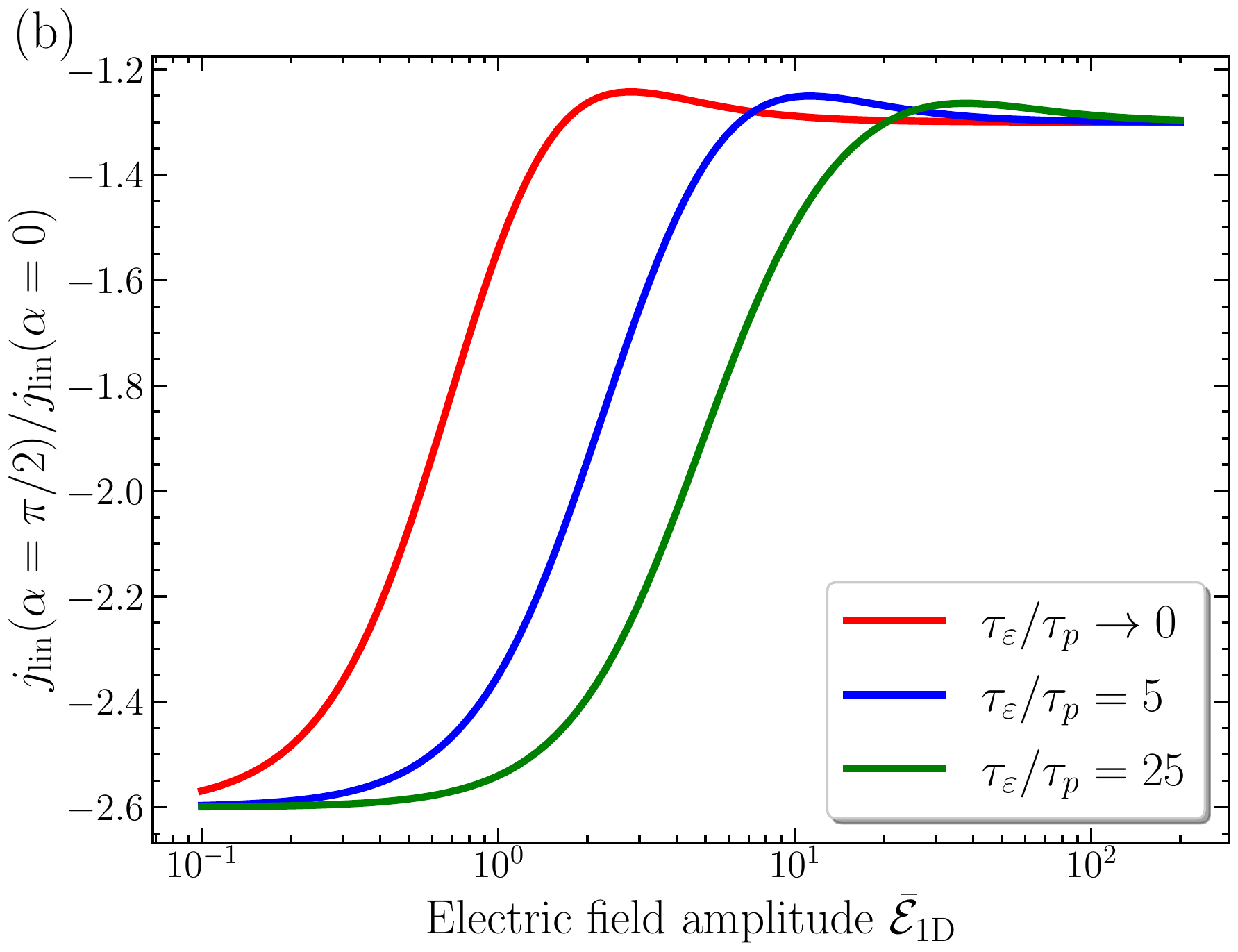} 
	\caption{Linear photocurrent dependencies on the orientation of the polarization  for $\tau_p/\tau_\varepsilon = 0.5$ (a) and the electric field amplitude (b) calculated after Eqs.~\eqref{j_lin1},~\eqref{j_lin_pol}. 
	The current in the panel~(a) is given in units 
	$j_1 v_0 \mu_{\rm B} g_1/(\omega D_2)$.
%%	$\overline{j}_0 \mathcal{C}_1$. 
%Here ${\cal E} = {\cal E}_{\text{circ}}$}. }
}
	\label{fig:lpge_angle}
\end{figure}

The photocurrent dependence on the orientation of the polarization plane is shown in Fig.~\ref{fig:lpge_angle} for $\theta = 0$.
In calculations we have taken $g_1/g_2 \approx 1.3$ relevant for HgTe-based quantum wells.
While in the linear in intensity regime 
the photocurrent values for the light polarization along and normal to the edge for $\theta= 0$ are related by
\begin{equation}\label{eq:j_lin_low}
	\frac{j_{1,\text{lin}}(\alpha = \pi/2)}{j_{1,\text{lin}}(\alpha = 0)}\biggr|_{I\to 0} = -\frac{D_2/D_1}{g_2/g_1},
	\end{equation}
at higher intensities this ratio depends on the electric field amplitude. In particular,   it depends on 
$\tau_\varepsilon/\tau_p$
at intermediate values of $\bar{\mathcal{E}}_{\text{1D}}$ and saturates in the high-intensity regime when $\bar{\mathcal E}_{\text{1D}} \gg 1$ 
at 
\begin{equation}\label{eq:j_lin_high}
	\frac{j_\text{lin}(\alpha = \pi/2)}{j_\text{lin}(\alpha = 0)}\biggr|_{I\to \infty} = -\frac{g_1}{g_2}.
\end{equation}

\section{Discussion}
\label{Disc}

The obtained results for absorption in 2D surface states are also relevant for nonlinear absorption in graphene.
Absorbance in the nonlinear regime was actively studied in graphene~\citep{Mishchenko2009,Agarwal2017,
Agarwal2018}. In the works~\cite{Agarwal2017,Agarwal2018} the density matrix formalism with the phenomenological relaxation rates for the population inversion $\gamma_1$ and for the interband coherence $\gamma_2$ was used. However, analysis shows that the absorbance in the model of Ref.~\cite{Agarwal2017} is governed by a single parameter, $\gamma_1\gamma_2$, and it can be obtained from the Eqs.~\eqref{eta_circ_1} and \eqref{eta_lin_1} by the substitutions $\tau_p \to \infty$, $\tau_\varepsilon\rightarrow 1/(2\sqrt{\gamma_1\gamma_2})$. By contrast, in the present work we demonstrate that the absorbance 
is not fully described by the time $\tau_\varepsilon$ but it is
also  sensitive to the efficiency of momentum relaxation processes. Figures~\ref{fig_LCD}--\ref{fig:abs_circ} and~\ref{fig:lin_dich} show that the absorbance in the nonlinear in intensity regime depends strongly on the one more kinetic parameter, $\tau_\varepsilon/\tau_p$.

Similarly, in
%This is also the case for 
Ref.~\cite{Matsyshyn2021} a nonlinear helicity-dependent photocurrent in 3D Weyl semimetals was calculated using  a combination of the non-equilibrium Green function technique and Floquet theory. 
However, the final result of Ref.~\cite{Matsyshyn2021}
%results for the photocurrent in the circular light polarization 
coincides with 
%the results 
that of 
kinetic approach~\cite{pssb_2019} similar to the present work,
%%our work
%Ref.~\cite{pssb_2019}  where we used the kinetic approach similar to the present work, 
in the limit $\tau_\varepsilon/\tau_p \to 0$.  
%However, as it follows from 
In this work we show for 1D edge states that the helicity-dependent photocurrent depends strongly on the ratio of the momentum and energy relaxation rates, see Fig.~\ref{fig:cpge}.
%and from Ref.~\cite{pssb_2019} for 3D systems
The same is true for the photocurrent in 1D edge states sensitive to the polarization plane orientation: Fig.~\ref{fig:lpge_angle} demonstrates that $j_\text{lin}$ also changes substantially with a variation of the relaxation times ratio.

%The photocurrent at intraband optical transitions in 3D topological semimetals~\cite{_EL_PRB_2018} is also 
%\commentNL{The paper~\cite{Dantas2021} is about "topological" nonlinearity for intraband transitions}

%In fact, our results shows that the inelastic relaxation modifies the absorption and photocurrent dependencies.

In the present work we use the simplest model of photocarrier energy relaxation described by a constant relaxation time $\tau_\varepsilon$ in Eqs.~\eqref{sys1}. Even in this model, the absorption and photocurrent behavior at high intensity can be more complicated due to accumulation of photoelectrons and photoholes in the states of the same band with smaller kinetic energy~\cite{Artemenko2013,ArtemenkoPRB}. This results in a change of the electron and hole chemical potentials, which can be taken into account in the intensity dependence of the factor $\mathcal F$, Eq.~\eqref{F}. In addition, $\mathcal F$ can change due to electron heating which results in an increased electron temperature in comparison with the temperature of the lattice~\cite{edge_nonlinear,BiTe_exp}.

%\commentNL{Say something about Artemenko?Possible:} \NL{Our equations~\eqref{j_j0_approx} shows, that if $\tau_\varepsilon \to \infty$ then the current is zero starting from some intensities. This situation could be treated properly taking into account $\tau_\varepsilon$ dependence from chemical potential similar to the work~\cite{Artemenko2013}.}.

Now we discuss the obtained results for 2D states.
For one-photon absorption, the analytical formula~\eqref{eta_circ_1} and Fig.~\ref{fig_LCD}(a) demonstrate that, in both polarizations, switching on the elastic scattering increases the absorbance for a fixed value of $\tau_\varepsilon$. This 
%happens because 
is explained by an emergence of an additional relaxation channel resulting in 
%a shortening of the total relaxation time $\tau$ and 
broadening of the energy stripe $\hbar/\tau$ for the final states which increases absorption.
The linear-circular dichroism value for one-photon absorption is not larger than unity for all intensities, Fig.~\ref{fig_LCD}(b).
This takes place because a linearly polarized light generates photocarriers anisotropically, see Eq.~\eqref{M_cv_quad} and inset to Fig.~\ref{fig_LCD}(a), i.e. an occupation of some final states is higher in comparison to the other. This leads to a faster saturation of optical transitions to these states and to a decrease of the absorbance in comparison with circular polarization where generation is isotropic. The elastic scattering 
%tends to make the photocarrier distribution isotropic. Therefore 
leads to isotropization of the photocarrier distribution. Therefore
the linear-circular dichroism disappears at $\tau_p \ll \tau_\varepsilon$, see Fig.~\ref{fig_LCD}(b) and Eqs.~\eqref{eta_circ_1},~\eqref{eta_lin_1}.

For two-photon absorption, the stronger is elastic scattering, the higher intensity is needed for emergence of the nonlinearity due to increase of $\hbar/\tau$. As a result, the linear-circular dichroism is the same as in the linear in intensity regime up to higher intensities at $\tau_\varepsilon/\tau_p \gg 1$, see Fig.~\ref{fig:2_phot_dich} and Eq.~\eqref{eq:ans}. At high intensity when $\bar{\mathcal E}_2 \geq 10$, the dichroism 
%degree saturates at the values which are higher 
is weaker for more efficient momentum scattering, Fig.~\ref{fig:2_phot_dich}.
%Similarly, for two-photon absorption, the stronger is elastic scattering, the smaller is the linear-circular absorption dichroism, see Fig.~\ref{fig:2_phot_dich}.
This is also explained by the effect of anisotropy in photogeneration for linear polarization which is not present for circular polarization, Eq.~\eqref{M_2_square}. Efficient elastic scattering leads to isotropization of the photocarrier distributions decreasing the effect of the difference in the generation rates for linear and circular polarizations. As a result, the absorption dichroism of two-photon absorption is suppressed. 

Absorption in 1D edge states also depends on the elastic relaxation efficiency, see Fig.~\ref{fig:abs_circ} and Eqs.~\eqref{abs_lin_pol},~\eqref{eq:lin_alp}. This is explained as in 2D case by the broadening of the energy space for the final states of the optical transition. This is also the reason for the dependence of the linear dichroism on $\tau_\varepsilon/\tau_p$ demonstrated in Fig.~\ref{fig:lin_dich}(b).

By contrast, the increase of the efficiency of elastic scattering leads to the decrease of the CPGE current, Fig.~\ref{fig:cpge}(a). It happens because, despite the absorption length increases and, hence, the photocurrent generation rate increases too, the relaxation time $\tau$ decreases with switching on an additional scattering channel. This effect prevails, therefore the higher is the elastic scattering rate, the smaller is the CPGE current, Fig.~\ref{fig:cpge}(b).
Comparing the circular photocurrent with its low-intensity values we see that the nonlinearity is weaker at higher momentum scattering rate, Fig.~\ref{fig:cpge}(a). This takes place because the back-scattered electrons leave the place in the momentum space where they are photogenerated and, thus, the saturation occurs at higher intensities.
The behavior of the photocurrent governed by the light linear polarization, $j_\text{lin}(\alpha)$, is also sensitive to the elastic scattering rate. In particular, it affects the ratio of its maximal and minimal values (realized at $\alpha =\pi/2$ and $\alpha=0$) at intermediate intensities, see Fig.~\ref{fig:lpge_angle}(b). The low and high intensity limits are given by Eqs.~\eqref{eq:j_lin_low} and~\eqref{eq:j_lin_high}, respectively.
%The nonlinear regime starts at higher intensities for higher momentum relaxation rates, and Fig.~\ref{fig:lpge_angle}(b).

%On the one hand, the increase of the efficiency of elastic scattering, thus the ratio $\tau_\varepsilon/\tau_p$, leads to the decrease of the current (Fig.~\ref{fig:cpge}(a)), that is in contrast to the absorption length (Fig.~\ref{fig:abs_circ}). It happens because we set $\tau_\varepsilon$ fixed, and increase of $\tau_\varepsilon/\tau_p$ leads to the decrease of the total $\tau$ that decreases the current.  On the other hand, the back-scattered electron leaves the place where the light generate electrons and, thus, the saturation of the optical transition, and, therefore the deviation of the current from the linear dependence~\eqref{j00}, happens at higher intensities for bigger value of $\tau_\varepsilon/\tau_p$.

Let us estimate characteristic intensities relevant for the nonlinear effects in absorption and photocurrents. We take the frequency $\omega/(2\pi)$ = 1 THz and the energy relaxation time $\tau_\varepsilon = 0.1$~ps~\cite{en_rel_Kvon}. For one-photon absorption in 2D states in topological insulators we define the saturation intensity $I_s$ by $\bar{\mathcal{E}}^2 = I/I_s$ with $\bar{\mathcal E}$ given by Eq.~\eqref{eq:bar_E_2D}. Taking $v_0 = 5\times 10^7$~cm/s~\cite{2D_params}  we get $I_s = 1.8$~kW/cm$^2$. 
%\NL{It can be seen from Figure~\ref{fig_LCD} that the LCD saturates at $ \bar{\mathcal{E}} \approx 10$ that, for chosen parameters, corresponds to $I \approx 18$~W/cm${}^2$ }
%
For two-photon absorption the saturation intensity $I_{s2}$ is defined by $\bar{\mathcal E}_{2} = I/I_{s2}$ with $\bar{\mathcal E}_{2}$ from Eq.~\eqref{E_2_circ_lin}. This yields for the same parameters $I_{s2}= 1.1$~kW/cm$^2$.
%
%have from Eq.~\eqref{E_2_circ_lin} \NL{$\bar{\mathcal E}_{2} = I/I_{s2}$, which yields $I_{s2}= 11 $~W/cm$^2$.}
%
For 1D systems we introduce the saturation intensity $I_{s1}$ via $\mathcal{E}_{\text{1D}}^2 = I/I_{s1}$. Estimation with $v_0 = 2.7 \times 10^7$~cm/s, $\abs{D_1/e} = 7\times 10^{-13}$ cm${}^2$, $D_2/D_1 = 2$~\cite{Durnev_JPCM} yields $I_{s1}= 3.5$~MW/cm$^2$.
The saturation intensity for 1D systems is much higher due to weakness of light absorption which occurs at forbidden optical transitions, and a higher  intensity is needed for its saturation.
Note that these values of laser intensity in the THz range are used in modern experiments~\cite{edge_nonlinear,BiTe_exp}.

\section{Conclusion}
\label{Concl}

We demonstrated that elastic scattering affects strongly nonlinear light absorption in topological insulators. For both one- and two-photon absorption at optical transitions between 2D surface states the saturation intensities and the linear-circular dichroism depend on the ratio $\tau_\varepsilon/\tau_p$. Similar effects take place in absorption in 1D edge states. The edge photocurrents generated in 1D states at both circularly and linearly polarized light of high intensity are also governed by elastic scattering efficiency. Estimates show that the studied effects can be detected in experiments. 
\\ 

\acknowledgments

The work  was supported by the Russian Science Foundation (Project~20-12-00147) and the Foundation for the Advancement of Theoretical Physics and Mathematics ``BASIS''.

\bibliography{Nonlin_TI}
\end{document}